# Luxury resale volume and composition shifted during the COVID-19 tourism shock, but co-purchase network tests were inconclusive and construction-sensitive

Tengfei Shao[1]*

[1] Global Education Center, Waseda University, Tokyo, Japan

*tengfei.shao@toki.waseda.jp

ORCID: 0000-0003-3089-5696

**Abstract**

Japan's 2020 tourism shock coincided with 27% fewer buyers and 35% lower gross merchandise value at one second-hand luxury resale intermediary. We compared 2019 and 2020 co-purchase projections of recorded brand labels with a Holm-corrected battery of five label-invariant metrics and Portrait Divergence against two permutation nulls rebuilding both graphs. Neither test separated the full years (smallest Holm p 0.407; 0.097 under the design-matched paired-swap null). For one random two-block split of brands, the battery detected a planted redirection of 30% of multi-brand baskets (178 of 10,399 second-group baskets, 1.7%) in 83 of 100 simulations (83–92 of 100 across 11 splits, all at least 80); Portrait Divergence reached 80% power only at full redirection. Observed falls in mean degree and clustering resembled those of 20–30% redirection, but the planted rise in degree assortativity was absent. A composition-only model of the observed attrition would be detected in only 15 of 100 simulations. Results moved with construction: without residual and material labels (post hoc), mean degree separated the years under the paired-swap null (Holm p = 0.030), and Portrait Divergence fell below 0.05, unadjusted, at the widest construction and within April–December. Shock-period network comparisons need rebuilt-graph nulls and a per-metric detection floor.

## Introduction

When an exogenous shock removes one demand channel from a market, the natural question is whether the transaction network that channel fed was restructured or merely thinned; on small thresholded projections, a distance or a scalar metric can move substantially through graph-reconstruction variance alone, so movement alone is not evidence of restructuring [1]. Two graphs built from different samples of one process differ by chance, and the size of that difference depends on how the graphs were constructed and on which properties a null model holds fixed [2,3]. Recent network studies of the COVID-19 period have read change from descriptive metric trajectories: a comparison of rail smart-card networks in 2019 and 2020 reported contraction and spatial reorganisation from classical and weighted indicators [4], and a study of urban mobility networks in 2020–2022 identified hubs and clusters without a pre-pandemic baseline [5]. Neither design states how large a change the comparison could have detected.

One resale intermediary's business-to-consumer records from 2018 to 2021 let us observe a buyer–brand co-purchase network before, during and after Japan's 2020 border closure. Until 2020, much of Japan's luxury retail demand was tied to inbound tourism and to purchasing on behalf of overseas buyers [6–9], and the pandemic halted international travel within weeks [10]. Second-hand luxury is a durable-value category whose buyers weigh consumption against retained value [11–16], and the pandemic's effects on retail and consumer spending have been documented in transaction-level data [17–21]. The closure thus supplies a large shock with a complete pre-shock baseline, treated here as a natural-experiment setting in the descriptive sense only: there is no control market.

Network-comparison statistics such as Portrait Divergence (PD) [22] summarise all-scale dissimilarity in one number and, in a benchmark on weighted networks, lost accuracy when the compared graphs differed in size and density [23]; null models tell us whether an observed value is unusual, but neither states what magnitude of change a non-detection could have missed, the quantity an informative null needs. The comparison literature offers many distances [24–28], evaluations of them [29] and a reframing of comparison as description [30]; temporal-network research treats a shock as a candidate change point [31,32], and planted-community detectability has a theory for block models [33]. A minimum detectable effect (MDE) under a stated alternative states which changes a non-detection is informative against. Two-sample tests for graphs have a power theory [34,35], network structure can be compared with formal tests [36], and permutation tests compare estimated networks [37], including symptom networks before and during the pandemic [38]; the shock-period network comparisons above report no MDE. Two earlier studies of the same firm's records used brand co-purchase networks: one compared the networks of four buyer groups with PD against random-partition nulls as a supplementary check and read each group's change across 2020 against a split-half noise floor [39], and one described the 2019-to-2020 PD of the annual network as the largest between-year break in the records [40]. Neither tested the full-market comparison against a null that rebuilds both graphs or stated a detection floor. Our earlier work applied network motifs to other second-hand luxury datasets [41–44]; this study differs in data, method and question.

Here we ask whether the 2020 shock restructured this intermediary's brand co-purchase network at community scale, treating a five-metric battery under two group-size-preserving permutation nulls as the primary family, and we make the detection floor explicit with a planted two-block alternative and add an empirical discrimination check on four exposure-cell networks; the protocol is then repeated on a public retail benchmark. Four questions organise the Results: how the shock registered in transaction volume and composition (RQ1); whether the 2019 and 2020 recorded-label projections differ on five label-invariant summaries beyond what the nulls produce (RQ2); what magnitude of community-scale change the battery could have detected, and whether the protocol behaves the same on a public benchmark (RQ3); and which buyer groups generate co-purchase pair mass, and how removing all cross-border buyers compares with removing equally many randomly drawn domestic buyers (RQ4). Supplementary Table S1 maps each claim to its evidence. Several results qualify the null and are reported as found: whether the years separated depended on graph construction and null, Portrait Divergence separated the years restricted to April–December, the battery rarely detected a composition-only attrition model, and a pre-specified size-matched removal showed the opposite sign.

## Results

### A concentrated network with shallow communities

The business-to-consumer (B2C) buyer–brand network over 2018–2021 comprises 42,962 buyers and 133 brand labels joined by 47,990 edges; a buyer links on average to 1.12 brands. Brand strength is highly concentrated (Gini 0.959; top-three share 0.769), as are sale values (Gini 0.825) and per-buyer spend (Gini 0.886): 6.5% of buyers account for 58.3% of gross merchandise value (GMV). Rolex (¥19.6 bn), Hermès (¥9.7 bn) and Patek Philippe (¥2.84 bn) lead the brand side, and one store accounts for ¥18.1 bn. Every annual co-purchase graph is disassortative (degree assortativity −0.55, −0.56, −0.56 and −0.45 for 2018–2021).

The pooled static projection (70 brands, 461 edges) has a three-community Louvain partition, watches (14 brand labels, ¥28.1 bn), fashion and leather (50 brand labels, ¥13.0 bn) and jewellery (6 brand labels, ¥0.47 bn), with weighted modularity $Q = 0.1381$, used here only as a descriptive partition quality (Supplementary Fig. S1). Against a degree-preserving null, the unweighted modularity is 0.1856 versus $0.1216 \pm 0.0064$ ($z = 10.04$ under the canonical node order, 5.6–5.9 under earlier orders; $p = 0.002$; Supplementary Fig. S2c), and every annual graph lies above its null (canonical $z = 4.18$, 4.72, 4.79 and 7.18; $p = 0.002$ each). Under every ordering tried, the static value exceeded the null maximum of 0.1411 and every annual value its null mean (empirical $p \leq 0.008$). The modal community count is 3, 2, 2 and 3 for 2018–2021, and the watch community has purity

1.000, 1.000, 1.000 and 0.875 (7 of 8 watch brands in 2021). The 2018 and 2019 partitions are less certain: Leiden–Louvain adjusted Rand index (ARI) 0.373 and 0.876 and mean pairwise ARI over 40 Louvain seeds 0.68 (minimum 0.37) and 0.70, against 1.000 and at least 0.97 later (Supplementary Fig. S2d).

### The shock was large in volume and composition

Annual B2C buyers numbered 12,802, 14,235, 10,399 and 9,220 in 2018–2021, and GMV was ¥11.6, 12.5, 8.1 and 9.5 bn (Fig. 1b,c). From 2019 to 2020, buyers fell 27% and GMV 35%. Composition moved in the tourism-exposed direction (Fig. 1d). The duty-free share of all transactions, business-to-business sales included, was 19.3% in 2018 and 22.1% in 2019, then 11.1% in 2020 and 7.5% in 2021; within offline retail, where duty-free purchasing concentrates, it fell from 67.2% of transactions in 2019 to 43.8% in 2020 and 26.3% in 2021. Duty-free sales were about twelve times as expensive as non-duty-free sales at the median (¥790,000 versus ¥65,455; Mann–Whitney U, two-sided $p < 0.001$; Fig. 1e). Cross-border buyers numbered 5,449 in 2019 and 2,016 in 2020 (Supplementary Table S5). The shock therefore coincided with a sharp fall, not a disappearance, in a high-value, tourism-linked slice of demand. Few buyers bridged the two years: 938 of the 14,235 buyers of 2019 were active in 2020 (9.0% of 2020 buyers). These changes are descriptive, and we relate them to the shock only in that they occurred during it.

### The primary tests did not separate 2019 and 2020

The primary construction, the recorded-label projection, keeps each year's 60 most frequent recorded brand labels, links two that share at least three buyers and removes isolates (Fig. 1a). It gives graphs of 40 nodes and 137 edges in 2019 and 33 nodes and 84 edges in 2020 (40 and 130 in 2018; 36 and 144 in 2021). The primary contrast is 2019 against 2020; 2018 and 2021 are descriptive context. We call the battery and PD at this construction the primary tests; the battery carries the inference and PD is a diagnostic. All five battery statistics and PD are unchanged when brand labels are permuted, so the tests compare graph shape, not which brands stay linked.

Under the buyer-year relabel null (R = 2,000 relabellings at the observed group sizes, 14,235 and 10,399), no metric of the five-metric battery separated the years (Table 1; Fig. 2b). The two largest moves, the clustering coefficient from 0.710 to 0.582 and mean degree from 6.85 to 5.09, had raw two-sided p = 0.099 and 0.081 and Holm-adjusted p = 0.407 each; the other three metrics had Holm p = 1.0, and the smallest Benjamini–Hochberg value was 0.25. PD between the two graphs was 0.673, inside the null (mean 0.541 ± 0.153, 95th percentile 0.754; one-sided p = 0.259; Fig. 2a).

Under the paired-swap null, which matches the sampling design (Methods), the battery again did not separate the years: mean degree had raw p = 0.019 and Holm p = 0.097, clustering Holm p = 0.372, and the other three Holm p = 1.0. PD had one-sided p = 0.118 (null mean 0.480, 95th percentile 0.721). This p-value is conditional on who was active in one year and who in both, and its exchangeability assumption (Methods) was not tested. On the union brand set, the two years' top-60 brands with isolates removed per year (41 and 33 nodes), no metric separated under the relabel null (Holm p 0.49–1.0; smallest raw p 0.098) and PD had p = 0.109.

We varied the construction in two directions (Table 2). At edge thresholds of two, three and five shared buyers the battery separated the years at no threshold. The resolution sweep crosses four constructions with the two nulls, giving eight PD tests. At the widest construction, top-111 with one shared buyer (87 and 70 nodes), PD had one-sided p = 0.0445 under the relabel null and 0.0225 under the paired-swap null, and both stayed below 0.05 under the alternative partial-sort rule for count ties at the rank cut-off (0.0460 and 0.0245; Supplementary Table S6). These values are nominal and unadjusted: Bonferroni gives 0.180 over the eight tests and 0.090 over the four paired-swap tests (Monte Carlo standard error of 0.0225, 0.0033). At top-80 with two shared buyers the paired-swap PD p was 0.173 under the brand-name tie rule and 0.0430 under the partial sort, so it depends on the tie rule. The other PD tests gave p = 0.118 to 0.322. The battery separated the years at no construction of the resolution sweep under either null or either tie rule; its smallest Holm p was 0.097, at the primary construction under the paired-swap null. The PD part of the null claim is therefore restricted to the primary construction.

Because 35.6% of 2020 transactions fell in January–March, before the closure took full effect, we repeated the primary comparison on April–December of each year (11,078 and 6,775 buyers; 35 nodes and 103 edges in 2019, 26 and 63 in 2020; relabel null, R = 2,000). The battery again did not separate the periods (Holm p = 1.0 for all five metrics), but PD did: observed 0.861, one-sided p = 0.019 (37 of 2,000 null draws at least as large). PD is a diagnostic outside the Holm family and this is one sensitivity test, but the PD part of the null claim holds for the full-year comparison only.

The recorded labels include two residual labels, "Other" and "No Brand", which are nodes of the primary graphs (Methods): "Other" has degree 17 in 2019 and 9 in 2020, and "No Brand" 5 and 4. Edges incident to these two labels contribute −0.37 of the −1.76 difference in mean degree, the other edges −1.39 (Supplementary Table S7). In a post hoc sensitivity analysis, specified after the primary results and the descriptive metrics of this construction had been seen, we removed them and four material labels before ranking and reran the comparison with the same units, group sizes and nulls (Supplementary Table S7). The named-brand graphs have 36 nodes and 115 edges in 2019 and 31 and 72 in 2020. Under the paired-swap null the battery separated the years through mean degree (6.39 to 4.65; Holm p = 0.030), and PD had one-sided p = 0.037; under the relabel null no metric separated (smallest Holm p 0.187) and PD had p = 0.098. Restricted to April–December, neither the battery (all Holm p = 1.0) nor PD (p = 0.062) separated the named-brand graphs. On this construction the whole procedure, under the relabel design, rejected in 3 of 100 unplanted replicates for the battery and for PD, and a planted two-block redirection of 30% of multi-brand baskets is detected with power 0.80 [0.708, 0.873]; the paired-swap procedure's size (battery 4 of 100, PD 5 of 100) is from the primary construction, not rerun here.

**What the battery could detect**

A non-detection is informative only if the test could have rejected a real change (Fig. 3).

The positive control is an empirical discrimination check under an adapted construction: the four exposure-cell sub-networks (Off-Tourist, Off-Local, On-Overseas and On-Local; 31–45 nodes) are built with the same projection steps and the cell rule of an earlier study of these records ([39]; Methods), and their buyer populations differ by construction, so this is an easier discrimination than the year comparison. The battery rejected all six cell pairs (smallest Holm p per pair 0.0025–0.0325; R = 2,000), whereas PD rejected two of six (Fig. 3b).

For the planted-effect power curve, each replicate split the 24,634 buyer-year baskets of 2019 and 2020 at random into groups of the observed sizes, 14,235 and 10,399, moved a fraction f of the second group's multi-brand baskets into a planted two-block brand split at fixed breadth, and tested the two groups through the same pipeline (100 replicates per f; Methods). At f = 0, the empirical size of the whole procedure, the battery rejected in 5 and PD in 2 of 100 replicates. Battery power was 0.10 [95% CI 0.049, 0.176], 0.33 [0.239, 0.431] and 0.83 [0.742, 0.898] at f = 0.1, 0.2 and 0.3, 0.97 at f = 0.4 and 1.00 from f = 0.5 (Fig. 3a; Supplementary Table S3a). For this random partition of the brands the battery MDE is therefore f = 0.3: at the observed group sizes, with between-sample variability, a planted two-block redirection of 30% of multi-brand baskets is detected with power 0.83 (multi-brand baskets: second-group baskets holding 2 to 30 of the pooled top-60 brands, about 593 per replicate; at f = 0.3, 178 of the 10,399 second-group baskets, 1.7%, were moved). At f = 0.3 the rejections came mainly from degree assortativity (rate 0.74) and clustering (0.63), then density (0.31), mean degree (0.30) and modularity (0.10). PD power was 0.17 [0.102, 0.258] at f = 0.3 and 0.66 [0.558, 0.752] at f = 0.8 and reached 0.8 only at f = 1.0, so its MDE is f = 1.0. Across the original and ten further random partitions (post hoc; Methods), the battery rejected in 83 to 92 of 100 replicates at f = 0.3 (median 85; 11 of 11 at least 0.8); under a partition by product category, a category-aligned planting that follows the watch-versus-other axis of the community core, it rejected in 51, 90 and 99 of 100 at f = 0.2, 0.3 and 0.4 (Supplementary Table S3e). Four of the five observed differences are of the size that planted redirection produces on average, density at 10–20%, clustering at about 20%, and mean degree and modularity at 20–30%, whereas degree assortativity is not. It carried most rejections at f = 0.3 (rejection rate 0.77 averaged over the random partitions, 0.83 for the category partition), and planting raised it by 0.08 at $f = 0.2$ and 0.18 at $f = 0.3$ on average (0.22 for the category partition at $f = 0.3$), but it changed by −0.005 from 2019 to 2020, below the no-planting mean of +0.007 (Supplementary Table S3f). With groups drawn and tested under the paired-swap null instead, the battery rejected in 4 and PD in 5 of 100 replicates at $f = 0$, and the battery detected the redirection at $f = 0.3$ with power 0.85 [0.765, 0.914].

A redirection is one alternative; the shock also changed who bought. In a composition-only model, the pooled baskets were relabelled so that the second group held the observed 2,016 cross-border and 8,383 domestic units, with within-type purchasing held at pooled behaviour (100 simulations; Methods). The battery detected this model in only 15 of 100 simulations (95% CI 0.086, 0.235) and PD in 8 of 100, so the comparison has little power against a change of this form. With the exposure labels assigned from each buyer-year's own sales instead of all four years (Methods), the battery detected it in 17 of 100 (0.102, 0.258). Descriptively, the observed 2019-to-2020 differences fell at percentiles 34, 10, 7, 97 and 46 of the simulated differences (density, mean degree, clustering, modularity, assortativity; Supplementary Table S3b).

The degree-preserving continuity test asks whether 2020 is closer to the real 2019 graph than to degree-preserving rewirings of it (2,000 rewirings; Fig. 3c); because PD ignores node labels, it asks whether 2020 shares structure of 2019 beyond the degree sequence, not whether the same brand pairs persist. Forward, PD(2020, real 2019) was 0.6734 against $0.7254 \pm 0.0229$ for rewired 2019 ($z = 2.27$, $p = 0.0155$); backward, $z = 0.95$ ($p = 0.179$). The forward signal appeared only at the three-shared-buyer threshold (Table 2b). Continuity is secondary evidence and, as the next paragraph shows, is not robust to simulated identity merging.

Buyer identity is a customer-name string; 0.7% of names are near-duplicates after normalisation, a measured rate that bears on splits, while the rate of exact homonyms, which would merge buyers, is unmeasured. We merged identities, split them, or mixed the two, at rates $\varepsilon$ from 0.007 to 0.20 (20 draws per cell; Fig. 3d; Supplementary Table S3d). The battery separated the years in 0/20 draws at $\varepsilon = 0.007$ in every mode (95% CI 0–0.168) and in at most 3/20 anywhere on the grid; PD rejected in at most 4/20. At $\varepsilon = 0.007$ the forward continuity signal remained significant in only 4/20 merge draws and 8/20 mixed draws (median z 0.58 and 1.44), but in at least 11/20 split draws up to $\varepsilon = 0.05$.

### The two-community core recurred in every year

On the brands common to each pair of years, the ARI between consensus partitions (100 Louvain seeds; co-assignment at least 0.5) was 0.433 for 2018–2019, 0.854 for 2019–2020, 0.456 for 2020–2021 (Supplementary Fig. S2a), 0.331 for 2018–2021, and each year's partition matched the pooled

static partition with ARI 0.35, 0.61, 0.72 and 0.74 (Supplementary Fig. S2b). The high 2019–2020 value partly reflects granularity: both consensus partitions have two communities, whereas 2018 and 2021 have three. We read the recurring watch-versus-leather core as a description, not as a test of structural change.

**Decomposition by exposure cell, reported as found**

The product menu was near-constant: 106, 111, 104 and 98 brand labels were offered in 2018–2021, 81 in all four years, and the annual top-60 sets overlapped with Jaccard 0.82–0.88 (48 brands in the top 60 every year; Fig. 4c). Only buyers of at least two brands create co-purchase edges: 3,315 buyers (7.7%) did so, and 3,239 bought at least two kept brands and so generate edges. Domestic buyers generated 64.4% of co-purchase pair mass and cross-border buyers 35.6% (Off-Local 47.1%, Off-Tourist 25.3%, On-Local 17.2%, On-Overseas 10.3%; Fig. 4a).

Removing all cross-border buyers from the pooled top-60 projection (58 nodes, 437 edges) left 56 nodes and 331 edges; 330 of the 437 original edges remained, an edge recall of 0.755 (Fig. 4b), and one edge is new because the re-ranked domestic top-60 differs by one brand. PD was 0.350 between the full and domestic-only graphs, and 0.720 and 0.707 between the cross-border-only graph and the full and domestic-only graphs; the menu Jaccard under removal was 0.967. Partition agreement under the same removal was low (co-membership preserved 0.518; seed-averaged ARI 0.052, maximum 0.401), but the domestic-only graph has weak community structure (weighted Q 0.052, against 0.131 for the full graph), so low agreement need not indicate reorganisation.

We pre-specified the expectation that removing cross-border buyers would cost fewer backbone edges than removing equally many randomly drawn domestic buyers. The sign was the opposite (Fig. 4d; Supplementary Table S5). Pooled, removing all 11,367 cross-border buyers retained 0.755 of the edges against a mean of 0.827 over 500 size-matched domestic draws, $\Delta = -0.071$ (2.5–97.5%: −0.124, −0.011; 6 of 500 draws above zero); $\Delta$ was −0.132 in 2018 (9/500) and −0.174 in 2019 (0/500), and −0.136 and −0.183 in the descriptive years 2020 and 2021. A diagnostic that matched the removed domestic buyers to the cross-border breadth distribution (D′) reversed the sign pooled ($\Delta' = +0.063$; 492/500 above zero; 2.5–97.5%: 0.006, 0.121) and in 2018 (+0.140); in 2020 (+0.033) and 2021 (+0.039) the mean was positive but the 2.5–97.5% range included zero. D′ was unavailable for 2019 because one domestic breadth stratum was smaller than its cross-border counterpart. D′ is a diagnostic, and neither contrast identifies a mechanism.

**The protocol transferred at an adapted threshold**

We repeated the protocol on UCI Online Retail II [45], the public transaction records of a UK online gift retailer (Fig. 5; Supplementary Table S4). After a pre-specified five-step cleaning (1,067,371 to 775,543 rows), two annual windows, W1 and W2, held 4,239 and 4,293 customers (2,708 in both), with the 60 most frequently purchased products as nodes. The literal thresholds of two, three and five shared customers produced complete graphs, so the pre-specified fallback rule selected $\tau^* = 93$, the smallest threshold of at least three giving W1 density at most 0.60 (W1: 1,042 edges, density 0.589; W2: 904 edges, density 0.511). This is an adapted-threshold transfer, on graphs denser and larger than the proprietary ones.

In 1,000 same-window half-splits, where no difference exists by construction, rejection rates were 0.027 [0.018, 0.039] for the battery and 0.043 [0.031, 0.057] for PD (Clopper–Pearson 95% CI): the battery does not over-reject, and its conservative size lowers its power. For the planted-effect curve, the 8,532 customer-window units were split at random into groups of 4,239 and 4,293, with the redirection planted in the second (100 replicates per f). At $f = 0$ the battery rejected in 2 and PD in 4 of 100 replicates. Battery power was 0.80 [0.708, 0.873] at $f = 0.2$ and 1.00 from $f = 0.3$ to 0.6; PD power was 0.19 at $f = 0.2$, 0.78 at $f = 0.3$ and 1.00 from $f = 0.45$. The battery MDE was 0.2 and the PD MDE 0.45, so the pre-specified two-part criterion (battery MDE below 1 and no larger than the PD MDE) was met. The benchmark's UK versus non-UK positive control was unevaluable (fewer than 20 nodes). A W1-versus-W2 comparison, which is not a shock contrast, gave relabel PD $p = 0.092$ and paired-swap PD $p = 0.033$, with no battery separation (smallest Holm p 0.670 and 0.255). The paired-swap null gave smaller PD p-values than the relabel null on both datasets; its size was measured on the proprietary data (battery 4 and PD 5 of 100 at $f = 0$), not on the benchmark.

**Discussion**

We asked whether a tourism shock restructured one resale intermediary's brand co-purchase network, measuring what a five-metric battery under two permutation nulls could detect. The 2020 shock coincided with a 27% fall in buyers and a 35% fall in gross merchandise value, while at the primary recorded-label construction the battery separated the full-year 2019 and 2020 graphs under neither null. There, the battery detected a planted two-block redirection of 30% of multi-brand baskets with power 0.83 for one random partition of the brands and 0.83 to 0.92 across 11, and it rejected all six pairs of exposure-cell networks, an easier discrimination. The observed falls in mean degree and clustering are of the size that planted redirection of 20–30% produces; the observed contrast lacks the rise in degree assortativity that carries most rejections under this planting mechanism, and the power analysis identifies neither the source of the observed declines nor changes of other forms.

Across analyses, the inference changed with graph construction and comparison window. At the recorded-label construction a composition-only model would have been detected in only 15 of 100 simulations (17 with unit-year labels), which leaves composition-driven change weakly probed. At top-111 with one shared buyer, PD was below 0.05 under both nulls and both tie rules, but the smallest p becomes 0.180 after Bonferroni correction across the eight sweep tests. Removing the residual and material labels post hoc gave a mean-degree battery separation under the paired-swap null (Holm $p = 0.030$; PD $p = 0.037$), whereas the relabel null gave none. Restricted to April–December, PD separated the recorded-label graphs ($p = 0.019$) but not the named-brand graphs ($p = 0.062$). Mean degree fell in every construction and window; the nulls also expect a fall, because the 2020 group is smaller (relabel mean −0.79 and paired-swap mean −0.45 at the primary construction), and at every construction of the resolution sweep the observed fall exceeded both null means without separating after Holm correction (Table 2a). The paired-swap procedure's size (battery 4/100, PD 5/100) is from the primary construction; the named-brand size read (3/100) uses the relabel design only. Exchangeability in the observed comparison is untested. These results support a construction-specific primary non-separation alongside secondary detections, without establishing shock-caused network reorganisation. In practice, the reading of this comparison depends on the construction and the null, and a test registered in advance on new data is the direct check.

The null concerns detectability, not identity, and it is relative to the two permutation references, whose sharp nulls are stated in Methods. The tests compare five label-invariant summaries of graph shape; which brand pairs stay linked lies outside them. The MDE is defined against a planted two-block redirection; a change of another form could be larger and still escape the battery, as the composition-only model shows. The positive control shows that the pipeline separates buyer populations with different brand mixes; it is not a ground truth, and the planted curve is the check with a known change. In a benchmark on weighted networks, PD lost accuracy when compared graphs differed in size and density [23], and our unweighted graphs have 40 and 33 nodes (35 and 26 for April–December). The nulls rebuild both graphs through one pipeline, so the size difference enters the null distribution, and the planted curve puts PD's MDE at $f = 1.0$. We therefore read PD as a diagnostic outside the Holm family and do not interpret its magnitudes.

The comparison with recent shock-period network studies is one of evidence form: they read change from descriptive indicators without a null model [4] or without a pre-pandemic baseline [5], whereas the present analysis adds a baseline year, two nulls and a detection floor, and its conclusion is correspondingly limited. The same holds for earlier readings of these records. One study read the offline cross-border cell network's change across 2020 (PD 0.756) against a split-half noise floor near 0.42 [39], and another described the full-market 2019-to-2020 PD of 0.67 as the largest between-year break in the records [40]. A split-half floor measures resampling noise, not the divergence that two groups of the observed, unequal sizes produce when the year label carries no information; against nulls that rebuild both graphs at those sizes, the full-market value lies inside both null distributions (relabel 95th percentile 0.754). The cell result concerns a different, smaller network and was not tested here. Null findings backed by a null network model [46] and proprietary network data with a configuration-model reference [47] have precedent in this journal.

The decomposition is reported as found. Domestic buyers generated 64.4% of pair mass, and removing every cross-border buyer left 75.5% of pooled backbone edges over a menu whose Jaccard stayed at 0.967. The pre-specified size-matched contrast went against our expectation: removing all cross-border buyers lowered edge retention more than removing equally many randomly drawn domestic buyers on every object of the pre-specified gate (pooled $\Delta = -0.071$; 6 of 500 draws above zero), and once breadth was matched (D′, diagnostic) the sign reversed pooled and in 2018. The pre-specified direction may have assumed too few multi-brand buyers among high-ticket cross-border buyers; D′ is consistent with this reading but was not designed to test it, and the decomposition identifies no mechanism.

What this case supports is a reporting discipline for small thresholded projections. A year-over-year PD of 0.67, read descriptively in earlier work as a structural break, lay inside both null distributions once both graphs were rebuilt at the observed group sizes; the battery's detection floor held across the brand partitions tested but rested mainly on one metric's response; and a composition-only change of the observed size was mostly undetectable. A metric move or graph distance on such projections should therefore come with a permutation null that rebuilds both graphs through the same pipeline, an MDE under the same correction read metric by metric, and the alternatives it does not cover.

The conclusions have explicit boundaries. The data come from one firm, one family of projection rules and annual snapshots of business-to-consumer sales, so sub-annual restructuring and market-wide behaviour lie outside what was tested; 35.6% of 2020 transactions fell before the closure took full effect. Buyer identity is a name string: the battery's non-separation is robust to identity noise, but the continuity signal remained significant in only 4/20 merge draws and 8/20 mixed draws at the simulated $\varepsilon = 0.007$, so continuity stays secondary. The named-brand analysis was specified after the primary results and is not a second primary test. A near-constant menu is necessary but not sufficient for a menu artefact; the positive control separates cell networks built on the same menu. The MDE refers to random splits of the pooled 2019 and 2020 baskets and to the specified breadth- and count-preserving redirection; it held across the 11 random brand partitions tested (power 0.83 to 0.92 at $f = 0.3$), other planting mechanisms were not tested, and the $f = 0.3$ interval of the original partition (0.742–0.898) contains 0.8, so the MDE follows the pre-specified point-power rule at grid resolution.

Three designs would sharpen these results: a persistent customer identifier, avoiding the merge problem for continuity; a transaction-level change-point design, which would reach sub-annual timing; and a second firm or period analysed with a construction registered in advance. The shock left a strong mark on volume and composition. At the recorded-label construction the tests did not distinguish the full-year co-purchase networks, although they detect a planted two-block redirection of 30% of multi-brand baskets with power 0.83 to 0.92 across the 11 random brand partitions tested; a post hoc named-brand construction distinguished the years under the paired-swap null, and a composition-shaped change of the observed size would mostly have escaped detection. The network comparison is therefore inconclusive and construction-sensitive, and it assigns no mechanism to any buyer group.

## Methods

### Study design and inference target

The study describes one firm's transaction network before, during and after Japan's 2020 border closure [7]. The closure followed entry restrictions that tightened progressively in early 2020, so the calendar-year 2020 window includes pre-closure months: 35.6% of 2020 B2C transactions (5,461 of 15,349) fell in January–March, and 38.1% of 2020 buyers made their earliest 2020 purchase in those months. The closure serves as a natural-experiment setting in the descriptive sense: the design has no control market and no difference-in-differences contrast, so volume and composition changes are reported as coinciding with the shock, not as effects identified from it. The inference target is whether the 2019 and 2020 annual recorded-label projections differ, on five label-invariant summary statistics (Supplementary Table S6), by more than two group-size-preserving permutation nulls produce; the tests do not assess whether particular brand pairs stay linked, and the partition comparisons that bear on this are descriptive. The primary contrast is 2019 against 2020; 2018 and 2021 are descriptive context.

### Data source, period and sample

The data are the sales records of one second-hand luxury resale intermediary in Japan for calendar years 2018–2021. Each record carries a buyer name string, a brand label, a channel (store or online), the online destination (domestic or overseas), a duty-free flag, a sale price and a date. We kept business-to-consumer sales with a positive price and non-missing buyer and brand fields. The brand field is the source record's brand name mapped to a standardised analysis label; five duplicate spellings of a brand (variant script or punctuation and one typo, each with at most five transactions) were mapped to that brand before any count, ranking or basket was built. The resulting buyer–brand network has 42,962 buyers, 133 brands and 47,990 edges. Annual buyers were 12,802, 14,235, 10,399 and 9,220 and annual GMV ¥11.6, 12.5, 8.1 and 9.5 bn. Of the 23,696 distinct buyers active in 2019 or 2020, 938 were active in both (6.6% of 2019 buyers, 9.0% of 2020 buyers, 4.0% of the union).

### Unit of analysis and target population

A buyer–brand bipartite graph links a buyer to each brand bought, and its one-mode brand projection links two brands by the number of buyers who bought both, the standard weighting of a bipartite projection [48]; projections of this kind underlie product networks [49] and co-purchase networks [50]. The unit of analysis is the annual brand co-purchase projection. The target population is this firm's B2C market, not the national market and not business-to-business sourcing.

### Variables and exposure cells

The five-metric battery comprises density, mean degree, average clustering coefficient, Louvain modularity and degree assortativity of the projection. PD [22] is the base-2 Jensen–Shannon divergence between the k-weighted distributions derived from the two graphs' network portraits, where the portrait counts, for each shortest-path length l, the nodes having k nodes at distance l; it is computed on the unweighted projection. Four exposure cells, which partition buyers without overlap, combine channel with a cross-border marker, as in an earlier study of these records [39]; a sale is cross-border when it is duty-free or an online sale to an overseas destination. Each buyer is assigned to one cell from all of the buyer's B2C sales in 2018–2021: offline if at least half of them are offline, otherwise online, and cross-border if at least half of them are cross-border, otherwise domestic. The cells are Off-Tourist (offline, cross-border), Off-Local (offline, domestic), On-Overseas (online, cross-border) and On-Local (online, domestic), and every purchase of a buyer counts toward that buyer's cell; 601 buyers (1.4%) had sales that on their own would fall in more than one cell (97 within 2019 and 67 within 2020). Because the cells are assigned from all four years, a unit can carry a different label from the one its own year's sales would give: applying the same rules to the sales of each 2019 or 2020 buyer-year alone changes the cross-border or domestic label of 18 of the 24,634 buyer-year units (0.07%; 10 of 14,235 in 2019 and 8 of 10,399 in 2020) and the cell of 176 (0.71%; 89 and 87). The decomposition and the size-matched removal therefore describe groups defined by the pooled labels. Buyers in Off-Tourist or On-Overseas are called cross-border and buyers in Off-Local or On-Local domestic. The duty-free flag marks tax-exempt sales, which include non-resident and export purchases, so it is an exposure marker rather than a tourist identifier; the cell name Off-Tourist is kept only as a label. Composition is measured by the duty-free share of sales, overall and within offline retail, and by the median sale price of duty-free and non-duty-free sales, compared with the two-sided Mann–Whitney U test.

### Buyer identity and pseudonymisation

The source records identify buyers by a customer-name string with no persistent verified identifier. The share of near-duplicate names after normalisation, 0.7%, is a measured rate; near-duplicates indicate possible splits (one buyer, two spellings), and this rate sets the lowest grid value, $\varepsilon = 0.007$. The rate of exact homonyms, which would merge two buyers, is unmeasured, so merge results are reported at simulated rates from $\varepsilon = 0.007$; the battery result holds up to $\varepsilon = 0.20$. In every data product that leaves the analysis environment, the name field is replaced by a salted, truncated cryptographic pseudonym.

**Time windows and the permutation unit**

Graphs are built per calendar year. No predictive model is trained, so there is no training–test split; the benchmark uses two annual windows and within-window half-splits (see below). The permutation unit is the buyer-year basket, the set of a buyer's purchases within one year. The 938 buyers active in both 2019 and 2020 contribute two baskets, and the two nulls treat these pairs differently (see Two permutation nulls).

**Network construction**

For each year we ranked brands by transaction count, kept the top 60, formed the weighted projection, removed edges with fewer than three shared buyers and removed isolated nodes. This primary construction was fixed in the analysis plan (see Pre-specification and exploratory analyses) and yields (nodes, edges) of (40, 130), (40, 137), (33, 84) and (36, 144) for 2018–2021. Its nodes are the recorded brand labels. The field holds six labels that are not named brands, the residual labels "Other" and "No Brand" and four material labels (K18 Gold, K24, Pt850 Platinum and Pt900 Platinum), and they enter the primary recorded-label construction and every construction derived from it; the named-brand sensitivity removes them before ranking (Supplementary Table S7). "Other" and "No Brand" are nodes of the primary graphs (ranks 10 and 20 by transaction count in 2019, 7 and 27 in 2020), whereas each material label has at most four transactions in 2018–2021. Ties at a rank cut-off are broken by ascending brand name, and one ranking function and one graph builder serve the observed graphs, every permutation draw and every planted replicate, so each observed statistic is the null statistic evaluated at the observed labels. Ties do occur (two brands share rank 60 in 2019 and three share ranks 59–61 in 2020), but none of the tied brands has an edge at the three-shared-buyer threshold, so the primary observed graphs do not depend on the tie rule. The pooled static projection used for description keeps the top 80 brands over 2018–2021 with the same threshold (70 nodes, 461 edges). A fixed shared-buyer threshold is one way to extract a projection backbone; statistically motivated filters are alternatives [51,52]. We kept a transparent fixed rule so that every null draw and planted alternative is rebuilt through the same pipeline, and we report threshold and resolution sweeps. Louvain was run with a fixed seed on graphs whose node and edge insertion order is canonical, which makes every reported partition deterministic. In the bipartite graph a buyer–brand edge carries the transaction count and the summed sale value; in the projection an edge carries the number of shared buyers. Modularity in the battery is the weighted Louvain modularity of this projection (0.177 in 2019 and 0.206 in 2020), and PD is computed from unweighted shortest-path lengths.

**Rationale for the inference protocol**

The analysis plan made the five-metric battery the primary family and PD a diagnostic; the planted curve, computed later, supports that order: PD's MDE at these graph sizes is $f = 1.0$, whereas the battery's is $f = 0.3$ for one random partition of the brands. Two nulls are used because each holds a different nuisance fixed: the relabel null holds both group sizes exactly, and the paired-swap null additionally keeps each returning buyer's two baskets together. The paired-swap null matches the sampling design, in which each returning buyer contributes one basket to each year; the relabel null, which can place both baskets of a returning buyer in one group, ignores this dependence and is reported alongside it, since both were specified before the runs. We report an MDE because a reader needs to know what a non-detection could have missed; without one a non-detection cannot be interpreted. Holm correction [53] controls the family-wise error rate over the five metrics within each null; the two nulls and the resolution sweep are not pooled into one family. The eight PD tests of the resolution sweep are diagnostics; they are reported unadjusted, with the Bonferroni value given for reference. With 100 replicates per f, the battery MDE is located at grid resolution by the frozen point-power rule: point power crosses 0.8 between $f = 0.2$ (0.33) and $f = 0.3$ (0.83). The Clopper–Pearson interval [54] at $f = 0.2$ (0.239–0.431) lies below 0.8, whereas the interval at $f = 0.3$ (0.742–0.898) contains 0.8.

**Primary family**

For each metric we computed the observed difference (2020 minus 2019) and its two-sided permutation p-value against $R = 2{,}000$ null draws, adjusted the five p-values with Holm's procedure and reported Benjamini–Hochberg values [55] as a secondary check. A metric separates the years when its Holm-adjusted p is below 0.05. The battery separates when at least one metric does. The two-sided p-value is $(b + 1)/(R + 1)$, where b counts null draws whose absolute difference is at least the absolute observed difference, so the test is centred on zero, not on the null mean. The original plan included six metrics. After the six-metric analysis, the sixth, the largest-connected-component fraction, which was 1.000 in both observed graphs and in every null draw, was removed, so the five-metric family is a post-result amendment. Under the original six-metric family the smallest Holm p is 0.489 under the relabel null and 0.117 under the paired-swap null, so neither family separates the years (Supplementary Table S6). Every battery result on the proprietary data was defined, with no non-finite null draws. PD was tested one-sided against the same draws, with $p = (b + 1)/(R + 1)$, where b counts null draws with PD at least as large as observed (at the primary construction under the relabel null, 518/2,001 = 0.259).

**Statistical reporting**

Tests use $\alpha = 0.05$. Metric differences are tested two-sided and PD one-sided. The Monte Carlo standard error of a permutation p-value is $\sqrt{(p(1 - p)/R)}$. Binomial proportions (power, size and identity-noise counts) carry Clopper–Pearson 95% intervals. A Mann–Whitney p-value below machine precision is reported as $p < 0.001$. Subsampling bands, year-over-year PD values, partition indices and the decomposition are descriptive and carry no significance claim.

**Two permutation nulls**

(a) Buyer-year relabel null. The 2019 and 2020 baskets were pooled, randomly relabelled into two groups of exactly 14,235 and 10,399 baskets, and both graphs were rebuilt through the full pipeline in each of R = 2,000 draws. A returning buyer's two baskets are relabelled independently; when both fall in one group they enter it as two separate buyers. The sharp null is that the year label carries no information about a basket's brands, so the 24,634 baskets are exchangeable between the years given the two group sizes. (b) Paired-swap null. Buyer-year baskets are the units; single-year baskets are relabelled with the per-group counts fixed, and the 938 both-year buyers swap their pair of baskets between the years with probability 1/2, so both groups again have exactly 14,235 and 10,399 baskets (R = 2,000). The pooled incidence reproduces the observed graphs. The paired-swap p-value is a conditional permutation p-value. Its validity requires exchangeability of the two buyer-year baskets within returning customers and exchangeability of single-year baskets conditional on group counts; this assumption was not tested. With groups drawn by the paired-swap scheme and no planted change (f = 0), the whole paired-swap procedure rejected in 4 of 100 replicates for the battery and 5 of 100 for PD (see Planted-effect power); like the relabel null's f = 0 row, this measures the size of the procedure under the null's own exchangeable scheme and so checks the implementation, not the exchangeability of the observed years. Its size was not measured on the public benchmark. A non-detection is therefore relative to these two permutation references, not to every form of structural change.

**Robustness constructions**

On the union brand set, where the union of the two years' top-60 brands is the candidate node set on both sides and isolates are removed per year (41 and 33 nodes), we reran the battery and PD under the relabel null. The threshold sweep rebuilt the top-60 graphs at thresholds t = 2, 3 and 5 shared buyers and ran the battery under the relabel null (R = 1,000). The resolution sweep crossed four constructions, top-60 with at least three shared buyers, top-80 with two, top-100 with two and top-111 with one, with both nulls (R = 2,000), giving eight PD tests; all eight are reported. Count ties at the rank cut-off occur in the observed graphs at top-60, top-80 and top-100, and only at top-80 do the two rules give different observed graphs (in 2019, 52 against 53 nodes); ties also occur in null draws at every construction, so the sweep was also run with the partial-sort tie-break in both observed and null graphs, on the same permutation draws, as a sensitivity analysis (Supplementary Table S6). The top-111 construction keeps every brand offered in 2019 and 2020 (111 and 104), so its observed graphs are identical under both rules and the two rules differ only in the null draws. The tie rule was fixed after the earlier sweep values had been seen (those values came from observed and null graphs ranked by two different sort routines and are recorded as superseded in Supplementary Table S6), so the sweep is reported under both rules; the reading rule, that a PD value whose two tie rules fall on opposite sides of 0.05 is called tie-dependent and is reported as nominal and unadjusted, was fixed before the rerun. A season-matched sensitivity compared April–December 2019 with April–December 2020, because 35.6% of 2020 transactions fall in January–March, before the border closure took full effect; it used the primary construction on buyer-period baskets and the relabel null (R = 2,000). The windows held 16,085 and 9,888 transactions from 11,078 and 6,775 buyers; no transaction lacked a date or had a date outside its source year.

**Named-brand sensitivity analysis**

After the primary results and the descriptive metrics of a construction without the non-brand labels had been seen, we wrote a protocol for a post hoc sensitivity analysis and fixed it before running it (Supplementary Table S6). The analysis removes the six non-brand labels from the pooled count matrix before top-60 ranking (in 2019 and 2020 only "Other", "No Brand", K18 Gold and Pt900 Platinum occur) and keeps the rest of the primary construction: the tie rule, the three-shared-buyer threshold, isolate removal and one graph builder. The permutation units are the primary ones, the 24,634 buyer-year baskets at group sizes 14,235 and 10,399, including 257 baskets that the filter leaves empty and that contribute no brand and no edge, and the paired-swap null uses the same 938 pairs. We ran both nulls (R = 2,000, the primary seed and streams), the April–December comparison on its buyer-period units (relabel null, R = 2,000) and a planted-effect power curve on this construction with the grid, replicates, inner R, size read, stop rule and MDE rule of the primary curve and an anchor set taken from the pooled named-brand top 60; both the primary curve and this size read use the relabel design only, without a paired-swap rerun. A parity run with no label removed reproduced the primary results exactly. Holm corrects within each battery, and the analysis is not a second primary test.

### Per-year uncertainty

For each metric and year we computed a Politis–Romano m-out-of-n subsampling interval [56] with $m = n/2$ buyers: the band is the observed value plus the deviation of the subsample 2.5th and 97.5th percentiles from the subsample mean, scaled by the without-replacement factor $\sqrt{m/(n-m)}$, which equals 1 at $m = n/2$. At this m the intervals are descriptive bands, not intervals with guaranteed coverage. All 20 metric-years of the five metrics (24 with the constant largest-component fraction) had 2,000 of 2,000 finite draws. Each band is centred on its point estimate, so it contains that estimate by construction of the centring, and this containment is not evidence of coverage. An n-out-of-n buyer bootstrap, which duplicates buyers and so inflates shared-buyer counts, is shown in Supplementary Table S2 only for contrast.

### Positive control

Each exposure-cell sub-network was built from the buyers of one cell over 2018–2021 with the same projection steps and the cell construction of an earlier study of these records [39], adapted to the smaller cells: the top 50 brands by summed sale value and at least two shared buyers (Off-Tourist 41 nodes, Off-Local 45, On-Overseas 31, On-Local 37). The battery and PD were applied to all six pairs with a relabel permutation that pools the two cells' buyers, relabels them at the observed group sizes and rebuilds both graphs by the same rule ($R = 2{,}000$). Observed PD between cells ranged from 0.45 to 0.81. A pair counts as rejected by the battery when any Holm-adjusted p is below 0.05. Because the cells share the firm's menu, the control asks whether the pipeline separates networks whose buyer populations differ while the product set is common; it is an empirical discrimination check, not a ground truth.

### Planted-effect power

Power was estimated at the real group sizes and with between-sample variability. Each replicate split the 24,634 buyer-year baskets of 2019 and 2020, the relabel test's own null population, at random and without replacement into groups of 14,235 and 10,399. A two-block redirection was planted in the second group only: the pooled top-60 brands were split once at random into two blocks of 30; a fraction f of that group's baskets holding 2 to 30 of these brands was selected; each selected basket kept its breadth and its transaction counts, its top-60 brands were mapped one-to-one onto brands drawn without replacement from one randomly chosen block, and its other brands stayed as they were. Both groups were then tested exactly as in the main test (same ranking, builder, battery and PD; inner relabel null $R = 999$), at $f \in \{0, 0.1, \ldots, 0.8, 1.0\}$ with 100 replicates per f. The $f = 0$ row is the empirical size of the whole procedure; no detection-floor claim was to be made if either test rejected in 11 or more of the 100 $f = 0$ replicates. The MDE is the smallest grid f at which power is at least 0.8 and the test is defined in at least 95% of replicates; power carries Clopper–Pearson 95% intervals. No unit held more than 30 anchor brands, and the second group held on average 593 eligible units at $f = 0$; at $f = 0.3$ on average 178 were moved, 1.7% of the second group's 10,399 baskets. The stop rule was not triggered (battery 5/100, PD 2/100 at $f = 0$). The composition-only alternative, without planting, relabelled the pooled baskets with the second group's cross-border count moved halfway ($a = 0.5$) or fully ($a = 1$) from its expectation under random splitting (3,151 of 10,399) to the observed 2,016, with 100 replicates each. The test was that of the planted curve (inner relabel null, $R = 999$). The per-metric rejection rates and, descriptively and without inference, the percentile of each observed 2019-to-2020 difference within the differences simulated at $a = 1$ are reported in Supplementary Table S3b. The paired-swap null was checked at three planted fractions. At $f \in \{0, 0.3, 0.4\}$, the groups were drawn by the paired-swap scheme, the redirection was planted in the second group as above and the inner null was the paired-swap null ($R = 999$; 100 replicates per point); the points were the battery MDE, its next grid value and $f = 0$. The anchor partition is fixed across replicates, so the curve is conditional on one random partition and on this planting mechanism. In a post hoc check, specified in a written protocol before it ran (Supplementary Table S6), the curve was recomputed at $f \in \{0.2, 0.3, 0.4\}$ for ten further random partitions of the same 60 anchors into two blocks of 30 and for one partition by product category, in which one block holds the 26 anchors whose modal product category is watch or accessory and the other the remaining 34 (100 replicates per f and partition; a mover chooses among the blocks large enough for its breadth). The cell seeds are those of the primary curve, so the groups, the movers and the inner draws are shared across the random partitions, which differ only in where movers are redirected; their range is therefore not an interval over partitions. For each partition the first tested f with power at least 0.8 is reported as a qualifying point on this restricted grid, not as a full-grid MDE. An earlier curve compared a fixed 2019 base with a planted copy of the same buyers; it had no between-sample variability and could not reject at $f = 0$, and it is kept in Supplementary Table S3c only as a superseded record. The corrected design was specified after that defect was found and before any of its results were seen (Supplementary Table S6).

**Degree-preserving continuity**

On canonical graphs, we compared PD(2020, real 2019) with the distribution of PD(2020, rewired 2019) over 2,000 degree-preserving double-edge-swap rewirings [3], each attempting 10m swaps within at most 200m tries on a graph with m edges, reporting the standardised distance z (real below rewired mean) and a permutation p-value, and repeated the test backward. A rewiring that exhausts the tries raises an error, and the test is then unevaluable. This belongs to the family of configuration-model references [2,47]. The threshold-sweep runs of the test (1,000 rewirings) are independent Monte Carlo replicates at $t = 3$. Continuity is secondary evidence.

**Identity-noise sensitivity**

We perturbed buyer identities at rate $\varepsilon \in \{0.007, 0.01, 0.02, 0.05, 0.10, 0.20\}$ in three modes: merge (a fraction $\varepsilon$ of buyers collapsed in pairs onto one identity), split (a fraction $\varepsilon$ of buyers whose rows are divided at random between two identities) and mixed ($\varepsilon/2$ of each), with 20 draws per mode and rate and one unperturbed run per mode. Each draw reran the battery and PD (relabel null, $R = 1,000$) and the forward continuity test (1,000 rewirings); counts carry exact 95% intervals. At $\varepsilon = 0.20$ under merging, 5 of 20 draws could not be rewired within the swap budget; their continuity tests are reported as unevaluable.

**Community detection and partition indices**

Louvain [57] maximised weighted modularity [58,59]; weighted Q is reported only as descriptive partition quality. Because modularity has a resolution limit and a degenerate optimum on small graphs [60–62], community significance was tested on the unweighted graph: unweighted Louvain Q against 500 degree-preserving rewirings per graph with the same swap budget [63], with z and an empirical p. The z magnitude depends on node order: we report z under the canonical order, and under every ordering tried the pooled static value lay above the null maximum and every annual value above its null mean (empirical $p \leq 0.008$). Consensus partitions pooled 100 Louvain seeds, linked brands co-assigned in at least half of the runs and took connected components [64]; consensus community counts were 3, 2, 2 and 3. Year-versus-year and year-versus-static ARI [65] were computed on consensus partitions over common brands, with a single seed-42 run as a verification row. Seed consistency is the mean pairwise ARI over 40 runs. Leiden partitions [66] were compared with Louvain by ARI. Watch purity is the share of watch brands in the watch-dominated community.

**Decomposition and size-matched removal**

Menu measures count the brands offered in each year's B2C sales and the Jaccard overlap of the annual top-60 sets; a buyer is edge-generating when it bought at least two kept brands. Pair mass is the number of shared-buyer brand pairs that a cell's buyers create among the kept brands, expressed as a share. Leave-cross-border-out rebuilt the pooled top-60 projection without cross-border buyers and reported edge recall (the share of the full graph's edges present after removal), PD, menu Jaccard, co-membership preservation and seed-averaged ARI. In the size-matched removal, ret_X is the fraction of an object's edges retained after removing all n_X cross-border buyers, and ret_D the same fraction after removing n_X randomly drawn domestic buyers (500 draws); $\Delta = \text{ret_X} - \text{ret_D}$. The pre-specified gate expected $\Delta > 0$ on its objects (pooled, 2018 and 2019); 2020 and 2021 are descriptive. The gate failed with the opposite sign on every object, and we report the result as found. Diagnostic D′ drew domestic buyers matched to the cross-border breadth strata of the full-object top-60; it is a diagnostic and was unavailable for 2019.

### Public-benchmark transfer

UCI Online Retail II [45], released under CC BY 4.0, holds the transactions of a UK online gift retailer. A pre-specified cleaning ledger removed cancelled or adjustment invoices, rows without a customer identifier, rows with non-positive quantity or price, non-product stock codes and exact duplicates (1,067,371 to 775,543 rows). Windows are W1, 2009-12-01 to 2010-11-30, and W2, 2010-12-01 to 2011-11-30. Customers play the role of buyers and the 60 products with the most rows per window that of brands; ties at rank 60 (about 39% of graphs) follow a frozen rule (count descending, then stock code ascending). The literal thresholds 2, 3 and 5 gave complete graphs, so the pre-specified rule chose the smallest $\tau \geq 3$ with W1 density at most 0.60, $\tau^* = 93$; we call this an adapted-threshold transfer. The size check drew 500 random customer half-splits per window and tested each with the relabel battery and PD ($R = 199$). Power was estimated as on the proprietary data: the 8,532 W1 and W2 customer-window units (customers active in both windows count twice) were split at random into groups of 4,239 and 4,293, a two-block redirection over the pooled top-60 products was planted in the second group, and both groups were tested with the frozen construction ($\tau^* = 93$; top-60 with the frozen tie rule) and an inner relabel null with $R = 199$, at the frozen grid $f \in \{0, 0.05, 0.1, 0.2, 0.3, 0.45, 0.6, 0.8, 1.0\}$ with 100 replicates per f. Per replicate, on average 3,183 of the 4,293 second-group units (74.2%) were eligible (2 to 30 anchor products) and 48 held more than 30 anchor products and were not redirected, against 593 of 10,399 (5.7%) eligible on the proprietary data; f is a share of eligible units, so equal f moves very different shares of all buyers and is not comparable across the two datasets. Undefined battery results, which arise when assortativity cannot be computed (2 of 100 replicates at $f = 0.8$ and all at $f = 1.0$), count as non-rejections. The pre-specified two-part criterion (battery MDE < 1 and battery MDE ≤ PD MDE) was applied to this curve; an earlier curve that compared a fixed base with a planted copy of itself is kept in Supplementary Table S4e as superseded. The UK versus non-UK positive control required at least 20 nodes per side and was unevaluable. W1 versus W2 was tested with both nulls ($R = 1{,}999$), with the 2,708 customers present in both windows as paired baskets; it is a non-shock comparison.

### Pre-specification and exploratory analyses

Pre-specified here means fixed in a dated internal written protocol before the analysis ran; no protocol was publicly registered. The primary construction and the battery were fixed in a written analysis plan in July 2026, after an initial comparison at this construction had found no separation;

the reported values are reruns with the corrected implementation. The benchmark gates (feasibility, size and power), the size-matched removal gate, the resolution sweep and its reporting branch, and the identity-noise grid were specified in dated protocols before the corresponding analyses ran. Supplementary Table S6 records every analysis with the date its specification was fixed, its criterion, its outcome and any deviation, including the change from six to five metrics; dated protocol copies are available to editors and reviewers. The corrected power designs (planted redirection at the real group sizes, the composition-only alternative, the season-matched comparison, the paired-swap check and the benchmark curve at split group sizes) were specified after the fixed-base defect was found and before any of their results were seen; the single graph builder, the tie rule and the merge of five duplicate brand spellings are implementation corrections made after the earlier results were seen, and the named-brand analysis is a post hoc sensitivity analysis (Supplementary Table S6). The planted curve under further block partitions, the composition-only alternative with year-specific exposure labels and the description of the residual labels are post hoc as well; their protocol was fixed before they ran. The benchmark gates passed; the size-matched removal gate failed. D′, the W1-versus-W2 comparison and the interpretation of the resolution sweep are exploratory.

**Ethics statement**

This study analyses archival commercial transaction records generated in the ordinary course of business by one resale intermediary and provided under a non-disclosure agreement that permits their research use in pseudonymised form; the firm is not named. The study involves no human participants, no intervention and no contact with individuals, and no data were collected from individuals for this research. Buyer names are replaced by non-invertible pseudonyms in every data product that leaves the analysis environment, and no personal identifier is published.

**Use of large language models**

Claude was used solely to polish the English language of this manuscript. All analysis and results are the author's own genuine work and were not generated by AI; the author takes full responsibility for the content.

**Software**

Analyses used Python 3.12.6 with NetworkX 3.2.1, NumPy 2.4.6, SciPy 1.17.1 and pandas 2.3.3; Leiden partitions used leidenalg 0.12.0 with python-igraph 1.0.0; partition indices used scikit-learn 1.9.0; figures used matplotlib 3.10.9; the benchmark workbook was read with openpyxl 3.1.5 and cached with pyarrow 24.0.0; network layouts in Supplementary Fig. S1 used ForceAtlas2 [67]. All random procedures draw from documented seeds (base seed 20260703), and the modularity null and continuity scripts reproduce their output byte for byte under different Python hash seeds.

## Code availability

The analysis code that produces the reported statistics runs on the confidential per-transaction records; it is available to editors and reviewers on request and, after publication, from the corresponding author (tengfei.shao@toki.waseda.jp) on reasonable request. A synthetic-data implementation of the construction, both nulls, the five-metric battery, the planted-effect power curve and the continuity test, together with the complete script for the UCI Online Retail II benchmark, is deposited on Zenodo (https://doi.org/10.5281/zenodo.21194626).

## Data Availability

The transaction data that support the findings of this study are available from a Japanese second-hand luxury resale firm, but restrictions apply to the availability of these data, which were used under a non-disclosure agreement for the current study and are not publicly available. The firm is not named publicly for commercial-confidentiality reasons; its identity has been disclosed to the editors confidentially. The data are available for peer review under a data-use agreement, a separate agreement between the firm and the reviewer, if requested by reviewers, and after publication from the corresponding author (tengfei.shao@toki.waseda.jp) upon reasonable request and with the permission of the firm. The public benchmark, UCI Online Retail II, is available at https://doi.org/10.24432/C5CG6D under a CC BY 4.0 licence.

**Acknowledgements**

The author thanks the resale firm for access to its transaction records under a non-disclosure agreement.

## Funding

This work was supported by MEXT Supporting Pioneering Research through AI for 1,000 Discovery challenges Program (SPReAD) Japan Grant Number JPMXP1726306059.

## Author contributions

T.S. is the sole author: Conceptualization, Methodology, Software, Validation, Formal analysis, Data curation, Visualization, Writing – original draft, Writing – review & editing.

## Additional information

**Competing interests:** The author declares no competing interests.

# Figures

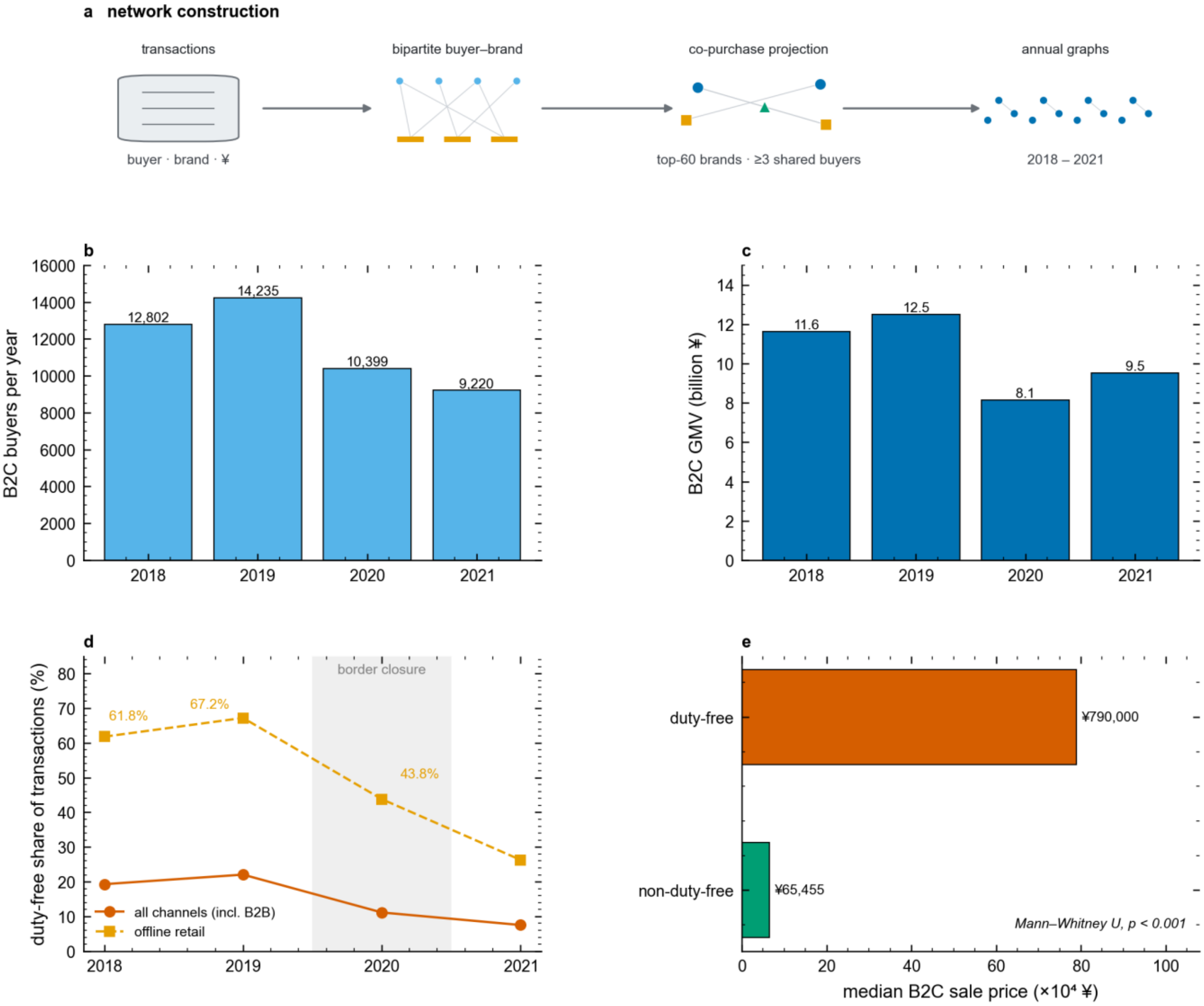

**Fig. 1. Network construction and the tourism shock.** (a) Construction pipeline. Business-to-consumer transaction records form a bipartite buyer–brand graph, which is projected to an annual one-mode brand co-purchase graph: the year's top-60 brands by transaction count, edges supported by at least three shared buyers, isolates removed. Every permutation null and planted alternative in this study is rebuilt through this pipeline; the positive control applies the same steps with a rule adapted to the smaller cell networks (Methods). (b) Annual B2C buyers, 2018–2021 (12,802; 14,235; 10,399; 9,220). (c) Annual gross merchandise value (¥11.6, 12.5, 8.1 and 9.5 bn). (d) Duty-free share of transactions across all channels, business-to-business sales included (19.3%, 22.1%, 11.1%, 7.5%), and within offline retail (61.8% in 2018, 67.2% in 2019, 43.8% in 2020, 26.3% in 2021); the shaded band marks 2020, the year of the border closure. (e) Median B2C sale price of duty-free and non-duty-free sales (¥790,000 versus ¥65,455; Mann–Whitney U, two-sided $p < 0.001$). Panels b and c are totals and panel d shares computed from the full records, not estimates. The pooled and annual co-purchase networks are drawn in Supplementary Fig. S1.

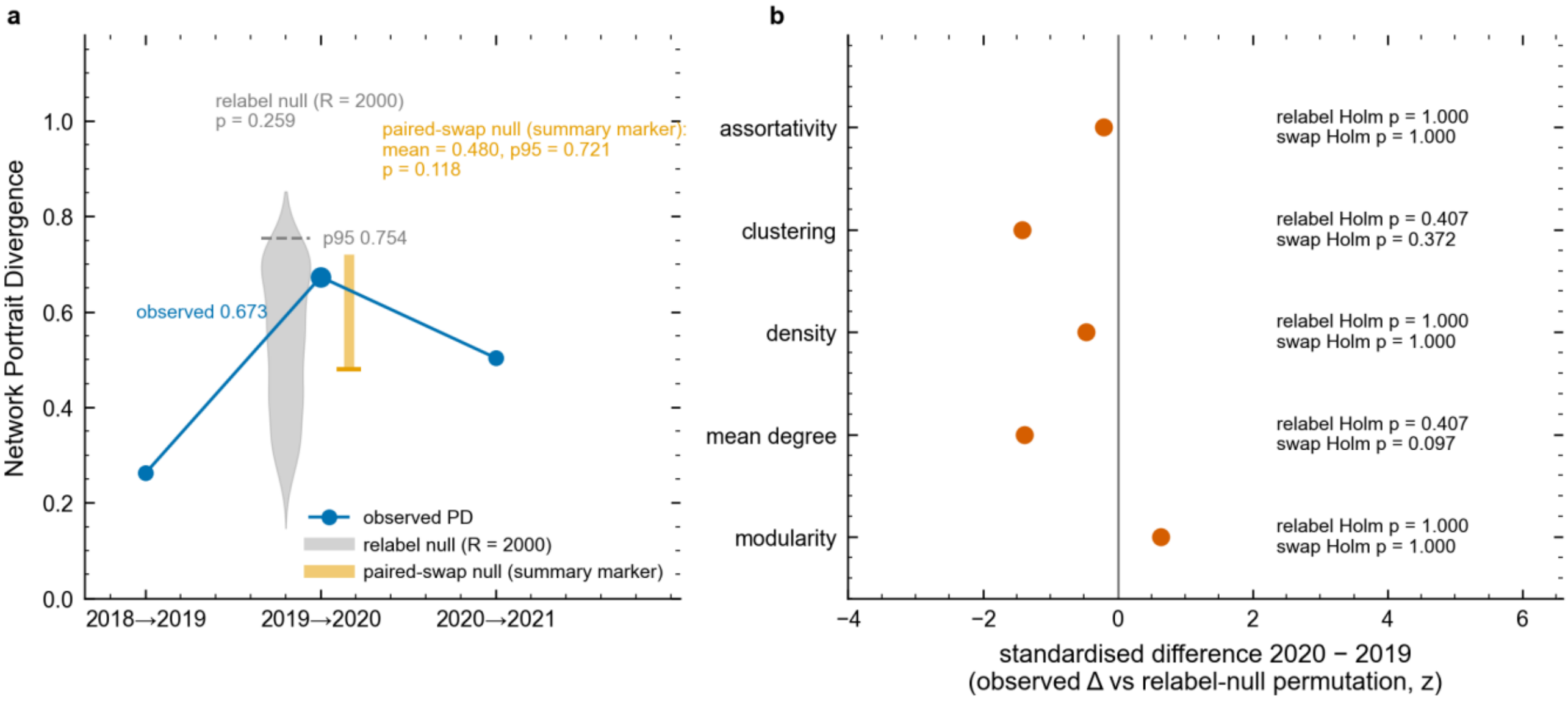

**Fig. 2. The structural null at the primary construction.** (a) Observed Portrait Divergence (PD) between the 2019 and 2020 graphs (0.673; 40 and 33 nodes) against the buyer-year relabel null (violin, R = 2,000; 95th percentile 0.754, one-sided p = 0.259) and the paired-swap null (bar from its mean, 0.480, to its 95th percentile, 0.721; one-sided p = 0.118; R = 2,000, per-draw values not stored). The line joining the three year-over-year values (0.262, 0.673, 0.503) is descriptive and not tested. (b) Standardised observed differences (2020 minus 2019) of the five metrics of the primary battery (density, mean degree, clustering coefficient, modularity, degree assortativity) relative to the relabel null distribution (z, centred on the null mean; the test statistic compares the absolute observed difference with absolute null differences, centred on zero; Methods), with Holm-adjusted two-sided p-values under both nulls annotated (smallest 0.407 under the relabel null and 0.097 under the paired-swap null). No metric separates the years under either null. The paired-swap p-values are conditional permutation p-values whose exchangeability assumption is stated in Methods and was not tested; the size of the paired-swap procedure on these data was 4/100 for the battery and 5/100 for PD at f = 0.

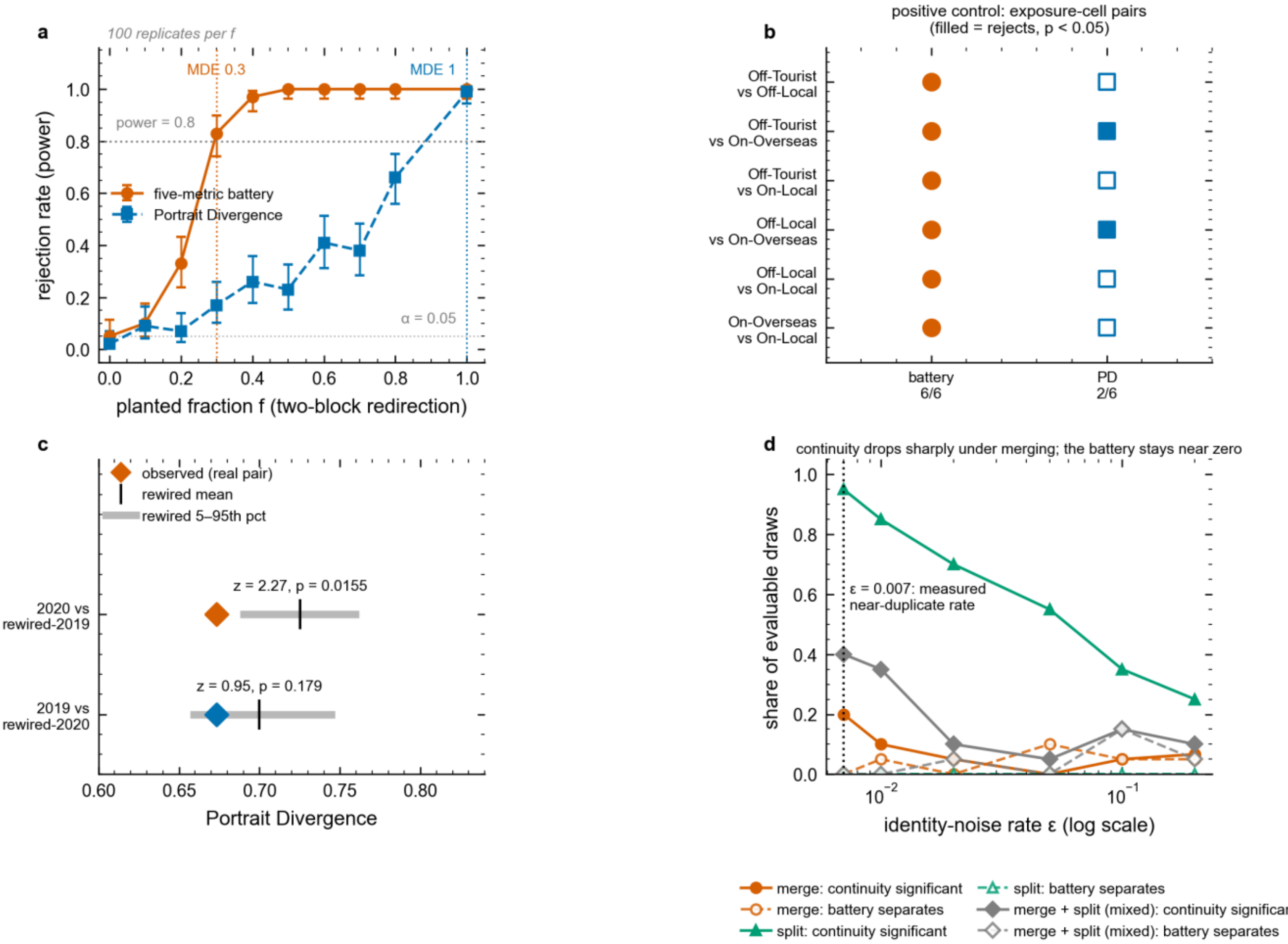

**Fig. 3. Detection floor, positive control, continuity and identity noise.** (a) Planted-effect power of the five-metric battery and of PD against the planted fraction f of multi-brand baskets moved into a two-block brand split, with the 2019 and 2020 baskets split at random into groups of the observed sizes (14,235 and 10,399) and the redirection planted in the second group (100 replicates per f; inner R = 999). Vertical bars are Clopper–Pearson 95% intervals; horizontal lines mark power 0.8 and α = 0.05. Size at f = 0: battery 0.05, PD 0.02. Battery power is 0.33 [0.239, 0.431] at f = 0.2 and 0.83 [0.742, 0.898] at f = 0.3, so the battery MDE is f = 0.3 at grid resolution for this random partition of the brands (0.83 to 0.92 at f = 0.3 across 11 partitions; Supplementary Table S3e); PD power is 0.66 [0.558, 0.752] at f = 0.8 and 0.99 at f = 1.0 (MDE f = 1.0). (b) Positive control on the four exposure-cell networks: filled markers denote rejection for a cell pair (battery 6/6, smallest Holm p per pair 0.0025–0.0325; PD 2/6; R = 2,000). (c) Degree-preserving continuity: observed PD of the real pair (diamond) against 2,000 degree-preserving rewirings of the reference year (band, 5th–95th percentile; tick, mean). Forward (2020 against rewired 2019): z = 2.27, p = 0.0155; backward: z = 0.95, p = 0.179. Continuity is secondary evidence and is not robust to identity merging at the simulated ε = 0.007. (d) Identity noise: share of evaluable draws (20 per cell; 15 for continuity under merging at ε = 0.20) in which the forward continuity test is significant and in which the battery separates the years, by rate ε and mode (merge, split, mixed); the dotted line marks ε = 0.007, the measured near-duplicate rate, at which merging is simulated. At ε = 0.007 continuity remains significant in 4/20 merge and 8/20 mixed draws, while the battery separates in 0/20 draws in every mode (95% CI 0–0.168).

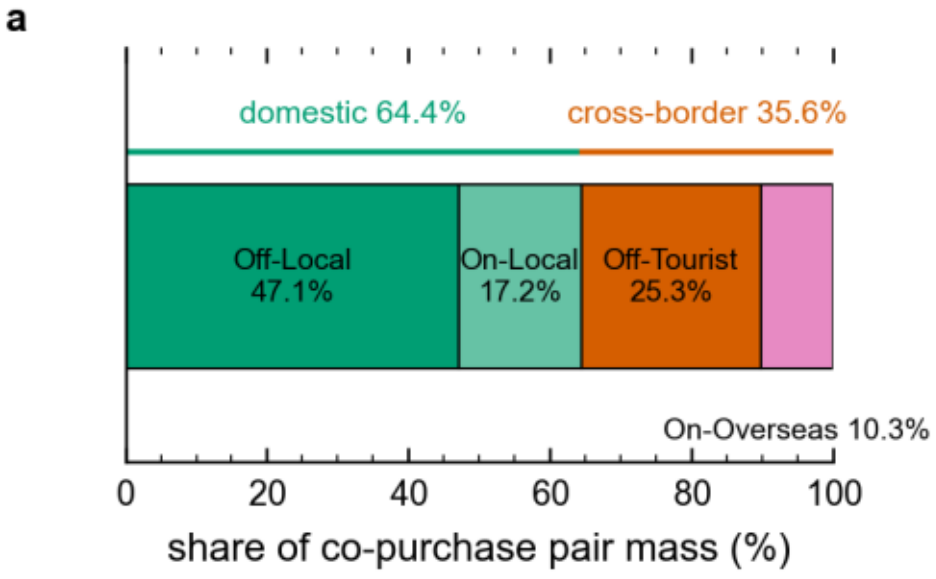

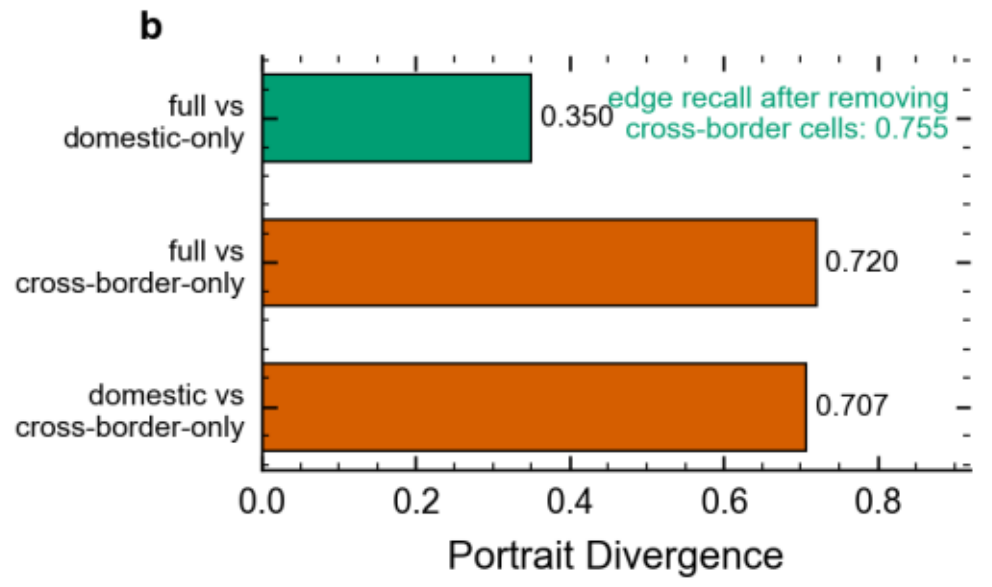

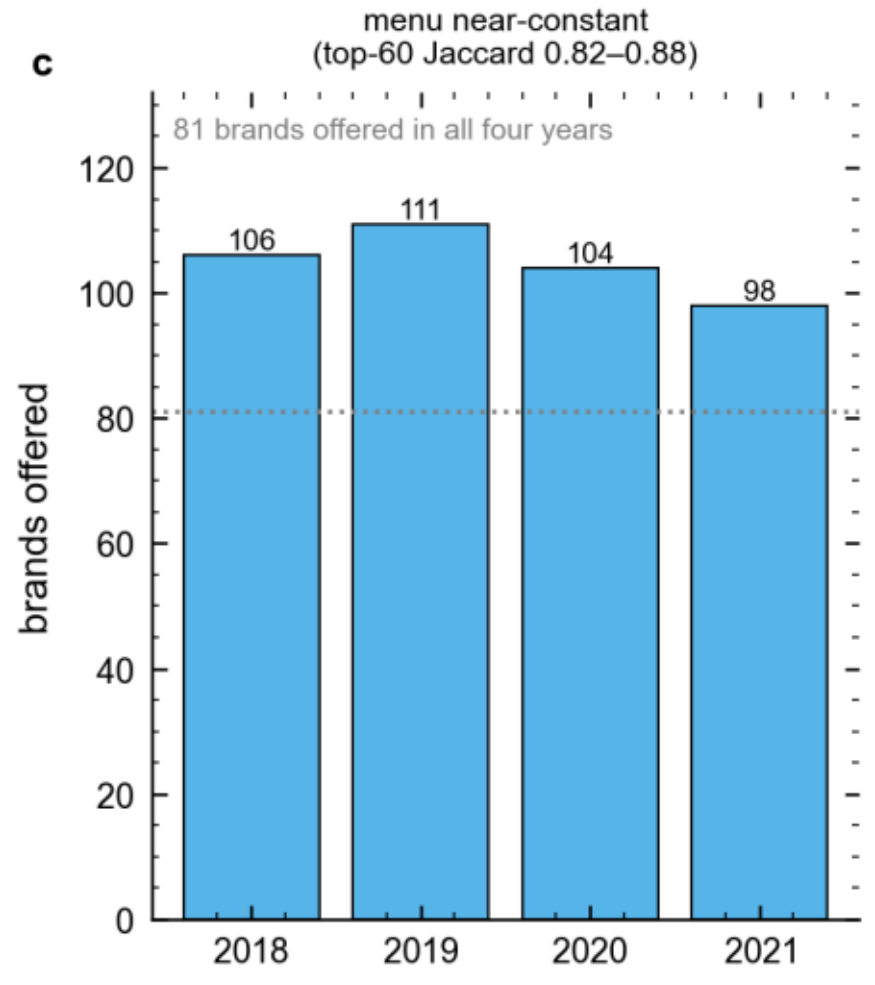

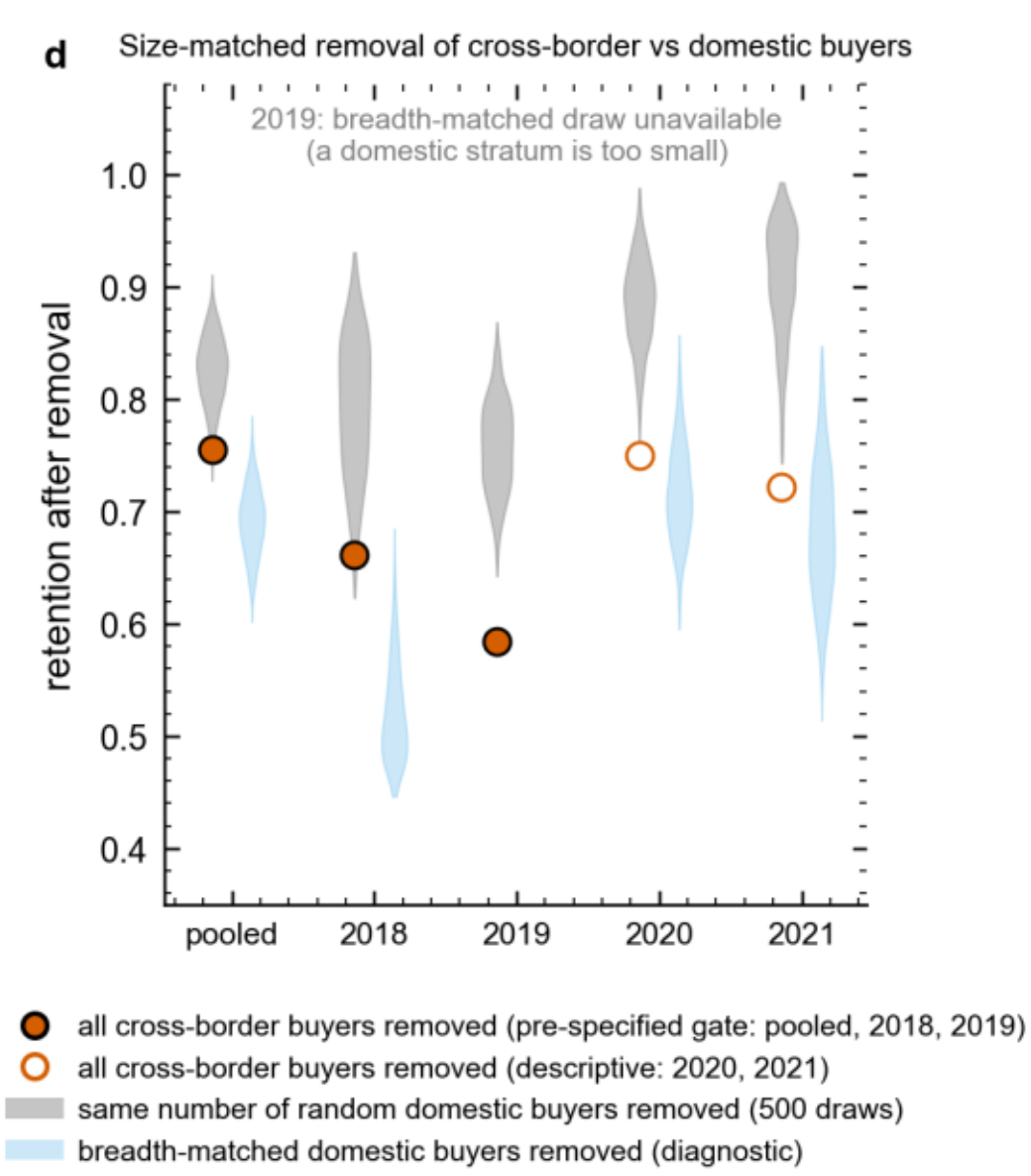

**Fig. 4. Decomposition by exposure cell.** (a) Share of co-purchase pair mass generated by each exposure cell: Off-Local 47.1%, Off-Tourist 25.3%, On-Local 17.2%, On-Overseas 10.3% (domestic 64.4%, cross-border 35.6%). (b) Leave-cross-border-out on the pooled top-60 projection (58 nodes, 437 edges): PD between the full and domestic-only graphs (0.350), between the full and cross-border-only graphs (0.720) and between the domestic-only and cross-border-only graphs (0.707); the domestic-only graph (56 nodes, 331 edges) keeps 75.5% of the full graph's edges. PD values are same-construction descriptions, not effect sizes. (c) Brands offered per year (106, 111, 104, 98); 81 brands are offered in all four years, 48 brands are in the top 60 every year, and annual top-60 sets overlap with Jaccard 0.82–0.88. (d) Size-matched removal: edge retention after removing all cross-border buyers (ret_X, marker) against the distribution of retention after removing equally many randomly drawn domestic buyers (ret_D, 500 draws; violin) for the objects of the pre-specified gate (pooled, 2018, 2019) and the descriptive years (2020, 2021). ret_X falls below ret_D on every object (pooled $\Delta = -0.071$; 6/500 draws above zero), the opposite of the pre-specified expectation. Breadth-matched domestic draws (D′, diagnostic) are shown as narrower violins alongside; D′ reverses the sign pooled ($\Delta' = +0.063$) and in 2018, and is unavailable for 2019.

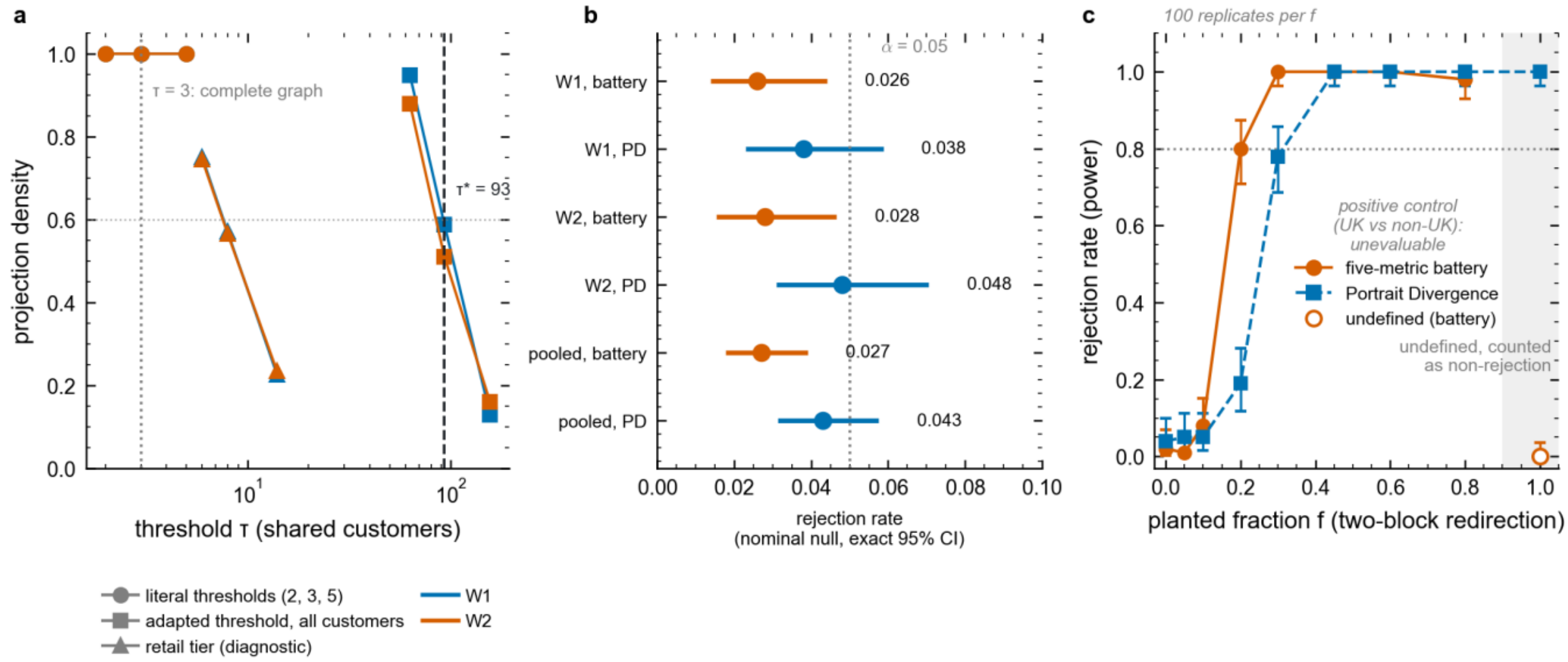

**Fig. 5. Public-benchmark transfer on UCI Online Retail II.** (a) Density of the W1 and W2 top-60 product co-purchase graphs against the shared-customer threshold τ; the literal thresholds 2, 3 and 5 give complete graphs, and the pre-specified fallback rule selects τ* = 93 (W1 density 0.589, W2 0.511). The third series is the retail-tier diagnostic construction (customers at or below the median total quantity; Supplementary Table S4b). (b) Rejection rates of the battery and PD in 1,000 same-window half-splits with Clopper–Pearson 95% intervals (battery 0.027 [0.018, 0.039]; PD 0.043 [0.031, 0.057]); the dashed line marks α = 0.05. (c) Planted-effect power at split group sizes: the 8,532 W1 and W2 customer-window units split at random into 4,239 and 4,293, the redirection planted in the second group (100 replicates per f; inner R = 199), with Clopper–Pearson 95% intervals. Battery MDE 0.2 (power 0.80 [0.708, 0.873]), PD MDE 0.45. At f = 1.0 degree assortativity cannot be computed on the planted graphs, so the battery point (open marker) is undefined and counted as a non-rejection. The superseded fixed-base curve is in Supplementary Table S4e. The UK versus non-UK positive control was unevaluable (fewer than 20 nodes) and is not shown.

## Tables

## Table 1. Five-metric battery and PD, 2019 versus 2020, primary construction (top-60 recorded brand labels, at least three shared buyers).

| Metric | 2019 | 2020 | Δ (2020 − 2019) | Relabel null Δ (mean ± sd) | Relabel raw p | Relabel Holm p | Paired-swap raw p | Paired-swap Holm p | Union-set Holm p |
|---|---|---|---|---|---|---|---|---|---|
| Density | 0.176 | 0.159 | −0.017 | −0.004 ± 0.028 | 0.550 | 1.000 | 0.539 | 1.000 | 1.000 |
| Mean degree | 6.85 | 5.09 | −1.76 | −0.79 ± 0.70 | 0.081 | 0.407 | 0.019 | 0.097 | 0.490 |
| Clustering coefficient | 0.710 | 0.582 | −0.128 | −0.024 ± 0.073 | 0.099 | 0.407 | 0.093 | 0.372 | 0.612 |
| Modularity | 0.177 | 0.206 | +0.029 | +0.004 ± 0.039 | 0.467 | 1.000 | 0.380 | 1.000 | 1.000 |
| Degree assortativity | −0.557 | −0.561 | −0.005 | +0.003 ± 0.035 | 0.896 | 1.000 | 0.901 | 1.000 | 1.000 |
| PD (diagnostic) | – | – | 0.673 | 0.541 ± 0.153 | 0.259 | – | 0.118 | – | 0.109 (raw) |

Notes: The five metrics form the primary family; p-values for them are two-sided with Holm adjustment over the five. Relabel null: buyer-year baskets pooled and relabelled into groups of exactly 14,235 and 10,399 (R = 2,000). Paired-swap null: the 938 buyers active in both years swap their pair of baskets with probability 1/2, single-year baskets are relabelled with group counts fixed (R = 2,000); these are conditional permutation p-values whose exchangeability assumption was not tested (size of the procedure on these data at f = 0: battery 4/100, PD 5/100; Methods). Union set: both years built on the union of the two years' top-60 brands, isolates removed per year (41 and 33 nodes), relabel null. PD row: observed PD in the Δ column, relabel null PD mean ± sd, and one-sided p-values; PD is a diagnostic outside the Holm family. The largest-connected-component fraction is 1.000 in both years and in every null draw and is excluded (Supplementary Table S6). The smallest Benjamini–Hochberg value under the relabel null is 0.25. Differences are computed at full precision and rounded once, so a displayed Δ can differ by 0.001 from the difference of the displayed values.

**Table 2. Construction sensitivity of the 2019 versus 2020 comparison.**
(a) Resolution sweep: eight PD tests (four constructions × two nulls; R = 2,000).

| Construction | Nodes 2019/2020 | Δ mean degree (2020 − 2019) | Null mean Δ mean degree, relabel / paired-swap | Observed PD | Relabel PD p | Paired-swap PD p | Smallest Holm p, relabel / paired-swap |
|---|---|---|---|---|---|---|---|
| Top-60, ≥3 shared buyers (primary) | 40/33 | −1.76 | −0.79 / −0.45 | 0.673 | 0.259 | 0.118 | 0.407 / 0.097 |
| Top-80, ≥2 | 52/43 | −1.36 | −0.70 / −0.39 | 0.671 | 0.281 | 0.173 | 0.502 / 0.500 |
| Top-100, ≥2 | 55/43 | −1.15 | −0.65 / −0.34 | 0.667 | 0.322 | 0.202 | 0.872 / 0.767 |
| Top-111, ≥1 | 87/70 | −1.30 | −1.07 / −0.51 | 0.649 | 0.0445 | 0.0225 | 1.000 / 1.000 |

(b) Threshold sweep at top-60 (battery under the relabel null, R = 1,000; forward continuity with 1,000 rewirings).

| Threshold t | Nodes 2019/2020 | Edges 2019/2020 | Observed PD | Metrics separating (Holm < 0.05) | Forward continuity z (p) |
|---|---|---|---|---|---|
| 2 | 47/41 | 197/135 | 0.604 | 0 | 1.38 (0.084) |
| 3 (primary) | 40/33 | 137/84 | 0.673 | 0 | 2.35 (0.015) |
| 5 | 30/24 | 88/49 | 0.833 | 0 | −1.23 (0.913) |

Notes: Δ mean degree and the null means are computed at full precision and rounded once; the nulls expect a fall because the 2020 group is smaller. PD p-values are one-sided. Ties at a rank cut-off are broken by ascending brand name in observed and null graphs. The top-111 PD p-values (shown to four decimals) are nominal and unadjusted across the eight construction-by-null tests: Bonferroni gives 0.180 over all eight and 0.090 over the four paired-swap constructions, and the Monte Carlo standard error of 0.0225 is 0.0033. Under the alternative partial-sort tie rule, applied to observed and null graphs on the same permutation draws (Supplementary Table S6), the relabel / paired-swap PD p-values are 0.261 / 0.119 (primary), 0.091 / 0.0430 (top-80, ≥2; 53 nodes in 2019, observed PD 0.733), 0.322 / 0.202 (top-100, ≥2) and 0.0460 / 0.0245 (top-111, ≥1), and the smallest Holm p is 0.095. The top-80 paired-swap value thus falls on the other side of 0.05 under the two rules, whereas both top-111 values stay below 0.05. The battery separates at no construction of the resolution sweep and no threshold. The t = 3 continuity row is an independent Monte Carlo replicate of the main test (z = 2.27, p = 0.0155, 2,000 rewirings).

## Supplementary information

The Supplementary Information file contains Supplementary Fig. S1 (the pooled and annual brand co-purchase networks drawn with a shared category encoding), Supplementary Fig. S2 (partition persistence: a, year-versus-year consensus ARI; b, year versus the pooled static partition; c, unweighted modularity against the degree-preserving null; d, seed consistency and Leiden–Louvain ARI), Supplementary Table S1 (claim-to-evidence map), Supplementary Table S2 (per-year five-metric values with m-out-of-n subsampling bands), Supplementary Table S3 (a, planted-effect power at the observed group sizes on the full f grid with Clopper–Pearson intervals and per-metric rates; b, the composition-only alternative, the season-matched comparison and the paired-swap check; c, the superseded fixed-base curve; d, the full identity-noise grid; e, planted-effect power under further partitions of the brands; f, observed against planted differences), Supplementary Table S4 (benchmark cleaning ledger, construction grid, size-check intervals, diagnostics and planted-effect power at split group sizes, with the superseded fixed-base curve), Supplementary Table S5 (size-matched removal and D′ by object) and Supplementary Table S7 (the post hoc named-brand sensitivity analysis: construction, both nulls, the April–December comparison and its planted-effect power). Supplementary Table S6 is the analysis record: each analysis with its status, the date its specification was fixed, its criterion, its outcome and any deviation.

Supplementary Information for: Luxury resale volume and composition shifted during the COVID-19 tourism shock, but co-purchase network tests were inconclusive and construction-sensitive

Tengfei Shao

This file contains the supplementary figures and tables referenced in the main text. Legends appear only here, next to each item. The primary family is the five-metric Holm battery (density, mean degree, clustering coefficient, modularity, degree assortativity) on the recorded-label projection; all five statistics are invariant to relabelling brands. Portrait Divergence is computed with the battery at the primary construction (together the primary tests) as a diagnostic outside the Holm family; the positive control, the planted-effect power curves, the degree-preserving continuity check, the partition indices and the post hoc named-brand sensitivity analysis (Table S7) are diagnostics.

## Fig. S1. Network construction and the annual co-purchase graphs

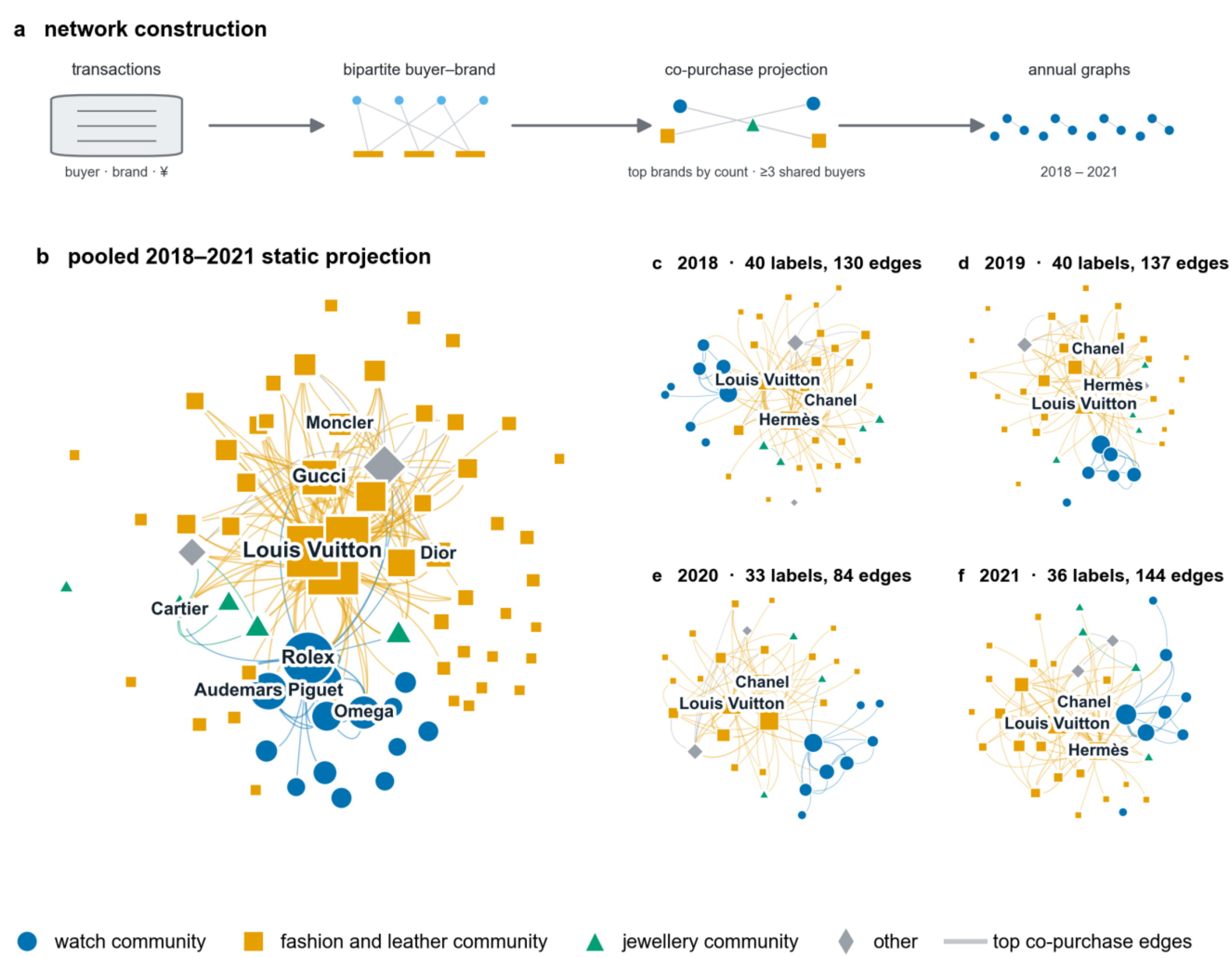

(a) Construction pipeline: buyer–brand transactions form a bipartite buyer–brand graph, which is projected to a brand–brand co-purchase graph among the year's top-60 brands by transaction count, with an edge placed between two brands when at least three buyers purchased both; isolated brands are removed; this rule is applied separately to each calendar year. (b) The pooled static projection, 2018–2021 (top 80 brands, 70 nodes, 461 edges), shown for description only and not used in any test. (c–f) The four annual graphs: 2018 (40 nodes, 130 edges), 2019 (40 nodes, 137 edges), 2020 (33 nodes, 84 edges), 2021 (36 nodes, 144 edges); only the 2019-to-2020 contrast (panels d and e) is tested, and 2018 and 2021 are descriptive context. Node colour and shape encode community labels, not each brand's own product type: each community of the pooled static partition is named after its dominant product type, as in the key: watch community (blue circles), fashion and leather community (orange squares) and jewellery community (green triangles); a brand whose own product type differs from that of its community carries the community's name. Grey diamonds (other) mark the residual brand labels "No Brand" and "Other", which are nodes of the recorded-label construction, and any annual-graph brand outside the pooled static projection. The static Louvain partition of panel b has three communities (weighted modularity Q = 0.1381); it is a descriptive summary and is not compared against a null model in this figure. Node size scales with the square root of weighted degree (summed shared-buyer counts); only the strongest edges by shared-buyer count are drawn (170 in panel b; 95 in 2018, 2019 and 2021; all 84 in 2020), with width proportional to that count. Each panel has its own ForceAtlas2 layout, so node positions are not comparable across panels.

## Fig. S2. Partition persistence

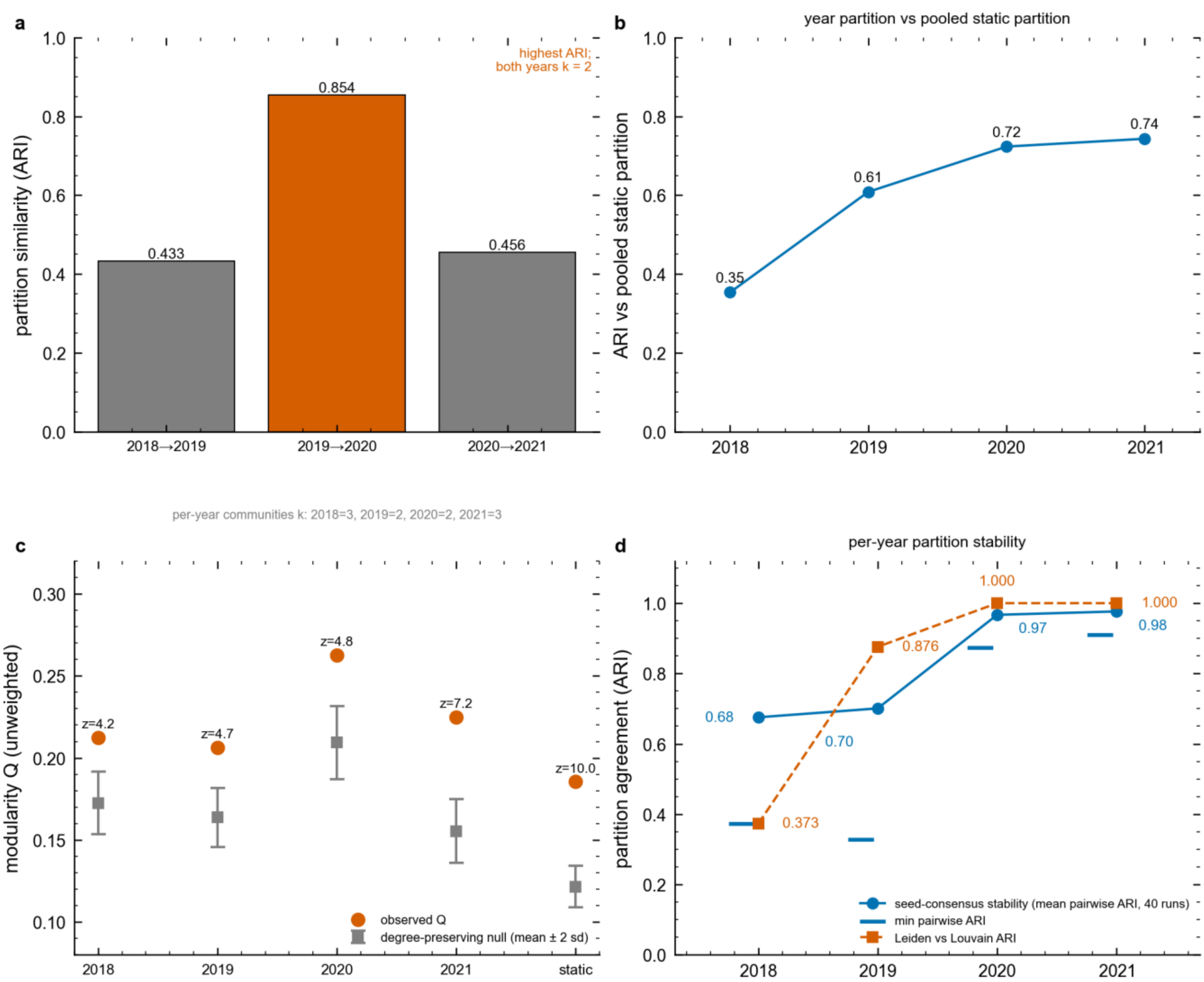

(a) Year-over-year Adjusted Rand Index (ARI) between consensus community partitions (100 Louvain seeds, co-assignment threshold 0.5, restricted to brands common to both years): 2018-to-2019 0.433, 2019-to-2020 0.854, 2020-to-2021 0.456. The 2019-to-2020 transition has the highest year-over-year ARI; this is partly a granularity coincidence, since both years' consensus partitions resolve to k = 2 communities (2019 individual seeds give k = 2 in 69 of 100 runs), while 2018 and 2021 resolve to k = 3. A single seed-42 run gives the same 2019-to-2020 value, 0.8538. (b) ARI of each year's consensus partition against the partition of the pooled static projection, restricted to common brands: 0.35 (2018), 0.61 (2019), 0.72 (2020), 0.74 (2021), rising monotonically. (c) Unweighted modularity Q_u of each year's graph and of the pooled static graph against a degree-preserving null (500 rewires per graph, mean plus or minus 2 standard deviations as error bars): every graph's observed Q_u exceeds its null (2018 z = 4.18, 2019 z = 4.72, 2020 z = 4.79, 2021 z = 7.18, static z = 10.04; all empirical p = 0.002). The z values shown use the canonical node order; they depend on the node-processing order used by Louvain (static z 5.56–5.85 and annual z 2.52–8.48 under earlier orderings). Under every ordering tried, the static Q_u exceeds its null maximum (0.141) and every annual Q_u exceeds its null mean (empirical p ≤ 0.008); under the earlier ordering the 2020 Q_u (0.2375) lay below that null's maximum (0.2484). This comparison is unweighted; the weighted modularity (Q = 0.138 for the pooled static graph) is not compared against this null. (d) Per-year partition stability: seed-consensus stability, the mean pairwise ARI across 40 Louvain runs per year with the minimum pairwise ARI marked separately, alongside the Leiden-versus-Louvain ARI. Consensus stability rises from 2018 to 2021 (mean 0.68, min 0.37 in 2018; mean 0.70, min 0.33 in 2019; 0.97 in 2020; 0.98 in 2021), and Leiden-versus-Louvain agreement follows the same pattern (0.373, 0.876, 1.000, 1.000). The low 2018–2019 values show the recovered partition is comparatively unstable to the Louvain seed and to the choice of community-detection algorithm in those years, consistent with the lower, more scattered ARIs for the same two years in panels (a) and (b).

## Table S1. Claim-to-evidence map

Primary = tested by the five-metric Holm battery at the primary construction, fixed in the analysis plan after an initial comparison and before the reported runs (Table S6). Diagnostic = informative but not part of the primary family. Descriptive = summary statistic, no test. RQ = research question (Introduction). R1–R7 = the seven Results subsections in order.

| # | Claim | Type | Evidence | Location | RQ |
|---|---|---|---|---|---|
| 1 | Buyers fell 27% and gross merchandise value fell 35% from 2019 to 2020 | Descriptive | 14,235 to 10,399 buyers; ¥12.5bn to ¥8.1bn | Abstract, R2, Discussion | RQ1 |
| 2 | Composition shifted: offline duty-free share fell from 67.2% to 43.8%; median price of duty-free sales ¥790,000 vs non-duty-free sales ¥65,455 | Descriptive with one test | Mann–Whitney two-sided $p < 0.001$ | R2 | RQ1 |
| 3 | No metric separates 2019 from 2020 at the primary construction under the buyer-year | Primary | Five-metric Holm battery, minimum Holm p = 0.407; R = 2,000 permutations | Abstract, R3, Table 1 | RQ2 |
|  | relabel null |  |  |  |  |
| 4 | No metric separates under the paired-swap null, which matches the sampling design | Primary (conditional permutation) | Minimum Holm p = 0.097 (mean degree); the paired-swap p-value's validity assumes exchangeability of returning-buyer basket pairs and of single-year baskets conditional on group counts, an assumption stated but not tested; the size of the whole procedure under this null on these data is 4/100 (battery) at f = 0 (Table S3b) | Table 1, Methods | RQ2 |
| 5 | Portrait Divergence (PD) is inside both null distributions at the primary construction in the full-year comparison | Diagnostic | Observed PD 0.673; relabel p = 0.259; paired-swap p = 0.118 | Abstract, R3, Fig. 2a | RQ2 |
| 6 | In the resolution sweep, PD p at top-111 with at least one shared buyer is 0.0445 (relabel) and 0.0225 (paired swap), and 0.0460 and 0.0245 under the alternative partial-sort tie rule: below 0.05 under both nulls and both tie rules, but nominal and unadjusted (Bonferroni 0.180 | Diagnostic (mandatory qualifier) | Four constructions times two nulls under two tie rules; battery separates at no construction | Abstract, R3, Table 2, Discussion | RQ2 |
| | over the eight tests and 0.090 over the four paired-swap tests; Monte Carlo standard error of 0.0225, 0.0033). At top-80, at least two shared buyers, the paired-swap PD p is 0.173 under the brand-name rule and 0.0430 under the partial sort, so it depends on the tie rule | | | | |
| 7 | The five-metric battery separates the two years at no construction in the resolution sweep | Diagnostic | Minimum Holm p across all eight construction-by-null cells = 0.097 (0.095 under the alternative tie rule) | R3, Table 2 | RQ2 |
| 8 | At the observed group sizes, with between-sample variability, a planted two-block redirection of 30% of multi-brand baskets is detected with power 0.83 [0.742, 0.898] for one random partition of the brands (100 replicates) and 0.83 to 0.92 across 11 random partitions; power at f = 0.2 is 0.33 [0.239, 0.431]; the empirical size at f = 0 is 5/100 | Diagnostic | Battery minimum detectable effect = 0.3 at grid resolution; per-metric rates in Table S3a; partitions in Table S3e | Abstract, R4, Fig. 3a | RQ3 |
| 9 | PD reaches power 0.8 only at f = 1.0; its power is 0.66 | Diagnostic | PD minimum detectable effect = 1.0 (Table S3a) | Abstract, R4, Fig. 3a | RQ3 |
| | [0.558, 0.752] at f = 0.8 and 0.17 [0.102, 0.258] at f = 0.3 | | | | |
| 10 | Empirical discrimination check on four exposure-cell sub-networks under an adapted construction (top 50 brands by summed sale value, at least two shared buyers): the battery rejects 6 of 6 pairs; PD rejects 2 of 6 | Diagnostic | Holm minimum p range 0.0025–0.0325 across pairs | R4, Discussion, Fig. 3b | RQ3 |
| 11 | Degree-preserving continuity signal, forward z = 2.27 (p = 0.0155), asymmetric (backward z = 0.95), significant only at threshold t = 3, not robust to identity merging at the simulated ε = 0.007 | Diagnostic (secondary) | 2,000 rewires; identity-noise grid (Table S3d) shows continuity significant in 4 of 20 merge draws and 8 of 20 mixed-mode draws at ε = 0.007 | R4, Discussion, Fig. 3c–d | RQ3 |
| 12 | Battery non-separation is robust to identity noise: 0 of 20 draws separate at ε = 0.007 in every perturbation mode; at most 3 of 20 separate anywhere on the grid | Diagnostic | Full grid in Table S3d | R4, Discussion | RQ3 |
| 13 | Every annual graph's community structure exceeds an unweighted degree-preserving null | Diagnostic | Under the canonical node order z = 4.18 to 7.18 across years and static z = 10.04, all empirical | R1, Fig. S2c | RQ2 |

|  |  |  | p = 0.002; under earlier orders annual z 2.52 to 8.48 and static z 5.56 to 5.85; under every ordering tried the static Q_u exceeds the null maximum and every annual Q_u its null mean (empirical p ≤ 0.008) |  |  |
|---|---|---|---|---|---|
| 14 | A two-community watch-versus-leather core recurs every year (watch-community purity 1.000, 1.000, 1.000, 0.875) | Descriptive | Per-year purity values | R1, R5 | RQ2 |
| 15 | The 2019-to-2020 transition has the highest year-over-year ARI (0.854) on consensus partitions, partly a granularity coincidence (both years resolve to k = 2) | Descriptive | Confirmed at a single seed (seed 42, value 0.8538) | R5, Fig. S2a | RQ2 |
| 16 | Domestic buyers generate 64.4% of pair mass; removing all cross-border buyers recalls 75.5% of edges; menu Jaccard 0.967; partition agreement after removal is low (co-membership preserved 0.518; seed-averaged ARI 0.052, maximum 0.401), and the | Descriptive | Pair mass by exposure cell; leave-cross-border-out network statistics | R6, Discussion, Fig. 4a–c | RQ4 |

| | domestic-only graph has weak community structure (weighted Q 0.052, against 0.131 for the full graph) | | | | |
|---|---|---|---|---|---|
| 17 | Removing all cross-border buyers lowers graph-level edge retention more than removing equally many randomly drawn domestic buyers (size-matched removal, a pre-specified gate that failed against its hypothesized direction): pooled $\Delta = \text{ret_X} - \text{mean ret_D} = -0.071$; in 6 of 500 draws, removing the domestic buyers retained fewer edges than removing the cross-border buyers ($\Delta = \text{ret_X} - \text{ret_D} > 0$) | Descriptive (pre-specified gate reported as found) | Per-object results for pooled and each year in Table S5 | R6, Discussion, Fig. 4d | RQ4 |
| 18 | Breadth-matched domestic draws ($D'$, diagnostic) reverse the sign of the graph-level edge-retention difference pooled and in 2018: pooled $\Delta' = +0.063$, with 492 of 500 draws exceeding zero; in 2020 and 2021 the | Diagnostic | Table S5 | R6, Discussion | RQ4 |
| | mean is positive but the 2.5–97.5% range includes zero; not available for 2019 | | | | |
| 19 | The inference protocol transfers to a public retail benchmark at an adapted threshold: size-controlled rejection rate 0.027 [0.018, 0.039] for the battery versus 0.043 [0.031, 0.057] for PD; at split group sizes (100 replicates per f; size at f = 0: battery 2/100, PD 4/100) the battery minimum detectable effect is 0.2 (power 0.80 [0.708, 0.873]) versus 0.45 for PD, so the pre-specified two-part criterion is met; battery results at f = 1.0 are undefined and counted as non-rejections | Diagnostic | Table S4c, S4e; Fig. 5c | R7, Fig. 5 | RQ3 |
| 20 | The brand menu is close to constant across the window (81 brands offered every year; 48 brands in the top 60 every year; top-60 Jaccard 0.82–0.88) | Descriptive | Necessary but not sufficient for the null: the positive control shows that the pipeline separates exposure-cell sub-networks built on the same menu | R6, Discussion, Fig. 4c | RQ4 |
| 21 | A rebuilt-graph permutation null and a stated minimum | Reasoned inference | This study's design, limited to small thresholded projections | Discussion | RQ3 |
| | detectable effect should accompany any reading of a metric move or graph distance as restructuring on a small thresholded network | | | | |
| 22 | A composition-only model of the observed attrition (the observed cross-border composition, within-type purchasing at pooled behaviour) was detected by the battery in only 15 of 100 simulations [0.086, 0.235] and by PD in 8 of 100, so the comparison has little power against a change of this form; with exposure labels assigned from each buyer-year's own sales, 17 of 100 [0.102, 0.258] | Diagnostic | Table S3b(i); observed differences at percentiles 34, 10, 7, 97 and 46 of the simulated differences (descriptive) | Abstract, R4, Discussion | RQ3 |
| 23 | Restricted to April–December of both years, the battery does not separate the periods (all Holm $p = 1.000$), whereas PD does (observed 0.861, one-sided $p = 0.019$); the PD part of the null holds for the full-year comparison only | Diagnostic (sensitivity) | Table S3b(ii) | Abstract, R3, Discussion | RQ2 |

| 24 | Under the paired-swap outer and inner design, the whole procedure rejects in 4/100 (battery) and 5/100 (PD) replicates at f = 0, and the battery detects the planted redirection at f = 0.3 with power 0.85 [0.765, 0.914] | Diagnostic | Table S3b(iii) | R4, Methods, Discussion | RQ3 |
|---|---|---|---|---|---|
| 25 | A post hoc named-brand construction (the residual labels "Other" and "No Brand" and four material labels removed before ranking) separates the full years under the paired-swap null through mean degree (Holm p = 0.030; PD one-sided p = 0.037) | Diagnostic (post hoc sensitivity analysis, not a primary test) | Table S7b | Abstract, R3, Discussion | RQ2 |
| 26 | Under the relabel null the named-brand construction does not separate the full years (smallest Holm p = 0.187; PD p = 0.098), and in April–December neither the battery (all Holm p = 1.000) nor PD (p = 0.062) separates the named-brand graphs | Diagnostic (post hoc sensitivity analysis) | Tables S7b, S7c | R3, Discussion | RQ2 |
| 27 | On the named-brand construction the whole procedure rejects in 3/100 replicates | Diagnostic | Table S7d | R3 | RQ3 |
| | (battery and PD) at f = 0, and a planted two-block redirection of 30% of multi-brand baskets is detected with power 0.80 [0.708, 0.873]; battery MDE 0.3, PD MDE 0.8 | | | | |
| 28 | Mean degree fell from 2019 to 2020 in every construction and window; the nulls also expect a fall, because the 2020 group is smaller, and at every construction of the resolution sweep the observed fall exceeded both null means without separating after Holm correction | Descriptive | Δ mean degree: primary −1.76 (Table 1); top-80 −1.36, top-100 −1.15, top-111 −1.30 (Table 2a); April–December −1.04 (Table S3b(ii)); named-brand full year −1.74 and April–December −1.16 (Table S7). Null means, relabel / paired swap: −0.79 / −0.45 (primary), −0.70 / −0.39, −0.65 / −0.34, −1.07 / −0.51 (Table 2a) | Discussion | RQ2 |
| 29 | The residual labels "Other" and "No Brand" are nodes of the primary graphs; edges incident to them contribute −0.37 of the −1.76 difference in mean degree | Descriptive | Degrees 17 and 5 in 2019, 9 and 4 in 2020 (Table S7) | R3 | RQ2 |
| 30 | Four of the five observed differences are of the size a planted redirection produces; degree | Descriptive | Table S3f | R4, Discussion | RQ3 |

| | assortativity, which carries most rejections at f = 0.3, fell slightly where planting raises it | | | | |
|---|---|---|---|---|---|

## Table S2. Per-year five-metric values with subsampling bands

Observed value of each of the five battery metrics for each annual graph, with its Politis–Romano *m*-out-of-*n* subsampling band: buyers are drawn without replacement at $m = n/2$ (2,000 draws per year), each metric is recomputed on the subsample, and the band is the observed value plus the deviation of the subsample 2.5th and 97.5th percentiles from the subsample mean, scaled by the without-replacement finite-population factor $\sqrt{(m/(n - m))}$, which equals 1 at $m = n/2$. The last column is the *n*-out-of-*n* buyer bootstrap percentile interval (2,000 draws with replacement), the method the band replaced; it is shown only for contrast. These bands are descriptive uncertainty summaries; they are not the permutation-null distributions used for the Holm battery in Table 1. Each band is centred on the full-sample point estimate and so contains it by construction of the centring; this containment is not evidence of coverage. Values are rounded half away from zero to three decimals. The largest-connected-component fraction is 1.000 in every year (all four annual graphs are connected); it was removed from the primary family after the six-metric analysis (Methods; Table S6) and is not listed here.

| Year | Metric | Observed | Subsample size *m* (of *n* buyers) | Politis–Romano *m*-out-of-*n* band (*m* = *n*/2), 95% | Band contains observed (by construction) | *n*-out-of-*n* bootstrap percentile interval, 95% (replaced method) |
|---|---|---|---|---|---|---|
| 2018 | density | 0.167 | 6,401 (of 12,802) | [0.132, 0.210] | yes | [0.128, 0.238] |
| 2018 | mean degree | 6.500 | 6,401 (of 12,802) | [5.575, 7.384] | yes | [5.381, 9.600] |
| 2018 | clustering coefficient | 0.703 | 6,401 (of 12,802) | [0.569, 0.825] | yes | [0.546, 0.741] |
| 2018 | modularity | 0.136 | 6,401 (of 12,802) | [0.090, 0.182] | yes | [0.124, 0.212] |
| 2018 | degree assortativity | −0.550 | 6,401 (of 12,802) | [−0.620, −0.469] | yes | [−0.537, −0.341] |
| 2019 | density | 0.176 | 7,118 (of 14,235) | [0.141, 0.217] | yes | [0.134, 0.212] |
| 2019 | mean degree | 6.850 | 7,118 (of | [6.076, 7.625] | yes | [5.957, 8.419] |

| | | | 14,235) | | | |
|---|---|---|---|---|---|---|
| 2019 | clustering coefficient | 0.710 | 7,118 (of 14,235) | [0.586, 0.831] | yes | [0.559, 0.750] |
| 2019 | modularity | 0.177 | 7,118 (of 14,235) | [0.125, 0.224] | yes | [0.151, 0.248] |
| 2019 | degree assortativity | −0.557 | 7,118 (of 14,235) | [−0.602, −0.504] | yes | [−0.499, −0.351] |
| 2020 | density | 0.159 | 5,200 (of 10,399) | [0.120, 0.210] | yes | [0.117, 0.198] |
| 2020 | mean degree | 5.091 | 5,200 (of 10,399) | [4.518, 5.715] | yes | [4.457, 6.865] |
| 2020 | clustering coefficient | 0.582 | 5,200 (of 10,399) | [0.446, 0.723] | yes | [0.460, 0.689] |
| 2020 | modularity | 0.206 | 5,200 (of 10,399) | [0.133, 0.270] | yes | [0.151, 0.274] |
| 2020 | degree assortativity | −0.561 | 5,200 (of 10,399) | [−0.646, −0.455] | yes | [−0.528, −0.338] |
| 2021 | density | 0.229 | 4,610 (of 9,220) | [0.186, 0.278] | yes | [0.142, 0.318] |
| 2021 | mean degree | 8.000 | 4,610 (of 9,220) | [6.879, 9.278] | yes | [5.833, 12.829] |
| 2021 | clustering coefficient | 0.752 | 4,610 (of 9,220) | [0.621, 0.879] | yes | [0.534, 0.784] |
| 2021 | modularity | 0.212 | 4,610 (of 9,220) | [0.152, 0.272] | yes | [0.172, 0.286] |
| 2021 | degree assortativity | −0.446 | 4,610 (of 9,220) | [−0.540, −0.340] | yes | [−0.455, −0.203] |

*Note: the band scales the subsample spread by the without-replacement finite-population factor √(m/(n − m)), which equals 1 at m = n/2; an earlier version scaled it by √(m/n), which made every band about 29% too narrow (Table S6). Each Politis–Romano band is centred on its observed value, so it contains that value by construction of the centring (20 of 20 metric-years); this containment is not evidence of coverage. The n-out-of-n buyer bootstrap, which duplicates buyers and so inflates shared-buyer counts, is the method the band replaced and is shown only for contrast; its interval misses the observed value for degree assortativity in 2018, 2019, 2020.*

## Table S3. Planted-effect power, alternative designs and identity noise

Tables S3a–c give the detection-floor analyses of the main comparison: the planted two-block redirection at the observed group sizes with between-sample variability (S3a; Design P), the composition-only alternative (Design S), the season-matched comparison and the paired-swap check (S3b), and the superseded fixed-base curve (S3c). Table S3e repeats the planted curve at f = 0.2 to 0.4 under ten further random partitions of the brands and one partition by product category, and Table S3f sets the observed differences beside the planted ones. Clopper–Pearson 95% intervals throughout. Table S3d is the identity-noise grid.

**Table S3a. Planted-effect power at the observed group sizes (14,235 and 10,399 buyer-year baskets per replicate; redirection planted in the second group; 100 replicates per f; inner relabel null R = 999).**

| f | Replicates | Battery rejections (power) [95% CI] | Battery defined share | PD rejections (power) [95% CI] | Density | Mean degree | Clustering | Modularity | Assortativity |
|---|---|---|---|---|---|---|---|---|---|
| 0 | 100 | 5/100 (0.05) [0.016, 0.113] | 1.00 | 2/100 (0.02) [0.002, 0.070] | 0.00 | 0.01 | 0.01 | 0.02 | 0.02 |
| 0.1 | 100 | 10/100 (0.10) [0.049, 0.176] | 1.00 | 9/100 (0.09) [0.042, 0.164] | 0.02 | 0.03 | 0.03 | 0.00 | 0.06 |
| 0.2 | 100 | 33/100 (0.33) [0.239, 0.431] | 1.00 | 7/100 (0.07) [0.029, 0.139] | 0.11 | 0.11 | 0.14 | 0.02 | 0.19 |
| 0.3 | 100 | 83/100 (0.83) [0.742, 0.898] | 1.00 | 17/100 (0.17) [0.102, 0.258] | 0.31 | 0.30 | 0.63 | 0.10 | 0.74 |
| 0.4 | 100 | 97/100 (0.97) [0.915, 0.994] | 1.00 | 26/100 (0.26) [0.177, 0.357] | 0.71 | 0.58 | 0.89 | 0.50 | 0.96 |
| 0.5 | 100 | 100/100 (1.00) [0.964, 1.000] | 1.00 | 23/100 (0.23) [0.152, 0.325] | 0.89 | 0.66 | 0.98 | 0.73 | 1.00 |
| 0.6 | 100 | 100/100 (1.00) [0.964, 1.000] | 1.00 | 41/100 (0.41) [0.313, 0.513] | 0.98 | 0.66 | 1.00 | 0.98 | 1.00 |
| 0.7 | 100 | 100/100 (1.00) [0.964, 1.000] | 1.00 | 38/100 (0.38) [0.285, 0.483] | 0.98 | 0.58 | 1.00 | 0.98 | 1.00 |
| 0.8 | 100 | 100/100 (1.00) [0.964, 1.000] | 1.00 | 66/100 (0.66) [0.558, 0.752] | 0.98 | 0.35 | 1.00 | 1.00 | 1.00 |
| 1.0 | 100 | 100/100 (1.00) [0.964, 1.000] | 1.00 | 99/100 (0.99) [0.946, 1.000] | 0.96 | 0.15 | 1.00 | 1.00 | 1.00 |

The per-metric columns are Holm rejection rates. The f = 0 row is the empirical size of the whole procedure (battery 5/100, PD 2/100; the pre-set stop rule, 11 or more of 100, was not triggered). The anchor brands are split into two blocks once, by the block seed 20260703; Table S3e gives further partitions. Battery MDE f = 0.3; PD MDE f = 1.0 (smallest grid f with power ≥ 0.8 and the test defined in ≥ 95% of replicates). The second group held on average 593 eligible units (2 to 30 of the 60 anchor brands); no unit held more than 30.

**Table S3b. Composition-only alternative, season-matched comparison and paired-swap check.**

(i) Composition-only alternative: the second group's cross-border count moved a fraction a of the way from its expectation under random splitting (3,151 of 10,399) to the observed 2,016; no planting; 100 simulations per a; inner relabel null R = 999. The rows marked unit-year labels repeat the design with each buyer-year's exposure label assigned from that year's sales alone (post hoc, protocol fixed before the run; Table S6): expectation 3,150 and observed 2,012 cross-border units in the second group.

| a | Replicates | Battery rejections (power) [95% CI] | Battery defined share | PD rejections (power) [95% CI] | Density | Mean degree | Clustering | Modularity | Assortativity |
|---|---|---|---|---|---|---|---|---|---|
| 0.5 | 100 | 7/100 (0.07) [0.029, 0.139] | 1.00 | 5/100 (0.05) [0.016, 0.113] | 0.02 | 0.01 | 0.00 | 0.04 | 0.02 |
| 1.0 | 100 | 15/100 (0.15) [0.086, 0.235] | 1.00 | 8/100 (0.08) [0.035, 0.152] | 0.01 | 0.02 | 0.00 | 0.12 | 0.01 |
| 0.5 (unit-year labels) | 100 | 3/100 (0.03) [0.006, 0.085] | 1.00 | 9/100 (0.09) [0.042, 0.164] | 0.00 | 0.00 | 0.00 | 0.03 | 0.00 |
| 1.0 (unit-year labels) | 100 | 17/100 (0.17) [0.102, 0.258] | 1.00 | 7/100 (0.07) [0.029, 0.139] | 0.01 | 0.03 | 0.00 | 0.15 | 0.02 |

Descriptive, without inference: the observed 2019-to-2020 difference of each metric and its percentile within the 100 differences simulated at a = 1: density −0.0166 (percentile 34); mean degree −1.7591 (percentile 10); clustering −0.1282 (percentile 7); modularity 0.0288 (percentile 97); assortativity −0.0047 (percentile 46).

(ii) Season-matched comparison, April–December 2019 against April–December 2020 (relabel null on buyer-period baskets, R = 2,000).

| Quantity | April–December 2019 | April–December 2020 |
|---|---|---|
| Transactions | 16,085 | 9,888 |
| Buyers | 11,078 | 6,775 |
| Nodes / edges | 35 / 103 | 26 / 63 |

| Metric | Observed difference | Raw p (two-sided) | Holm p |
|---|---|---|---|
| density | 0.021 | 0.507 | 1.000 |
| mean degree | −1.040 | 0.452 | 1.000 |
| clustering | 0.022 | 0.870 | 1.000 |
| modularity | 0.023 | 0.680 | 1.000 |
| assortativity | 0.030 | 0.583 | 1.000 |
| PD (diagnostic) | 0.861 (observed) | 0.019 (one-sided; 37 of 2,000 draws ≥ observed) | – |

Excluded transactions: 0 with a missing date, 0 with a date outside the source year.

(iii) Paired-swap check: groups drawn by the paired-swap scheme, redirection planted in the second group, inner paired-swap null R = 999, 100 replicates per f.

| f | Replicates | Battery rejections (power) [95% CI] | Battery defined share | PD rejections (power) [95% CI] | Density | Mean degree | Clustering | Modularity | Assortativity |
|---|---|---|---|---|---|---|---|---|---|
| 0 | 100 | 4/100 (0.04) [0.011, 0.099] | 1.00 | 5/100 (0.05) [0.016, 0.113] | 0.01 | 0.01 | 0.01 | 0.01 | 0.01 |
| 0.3 | 100 | 85/100 (0.85) [0.765, 0.914] | 1.00 | 8/100 (0.08) [0.035, 0.152] | 0.28 | 0.26 | 0.54 | 0.12 | 0.74 |
| 0.4 | 100 | 99/100 (0.99) [0.946, 1.000] | 1.00 | 20/100 (0.20) [0.127, 0.292] | 0.63 | 0.50 | 0.86 | 0.44 | 0.98 |

**Table S3c. Superseded (fixed-base planted contrast; no between-sample variability): the earlier planted-effect curve, a fixed 2019 base against a planted copy of the same buyers at equal group sizes (24 replicates per f; R = 1,000).**

| f | Battery rejections (power) [95% CI] | PD rejections (power) [95% CI] |
|---|---|---|
| 0 | 0/24 (0.00) [0.000, 0.143] | 0/24 (0.00) [0.000, 0.143] |
| 0.1 | 0/24 (0.00) [0.000, 0.143] | 0/24 (0.00) [0.000, 0.143] |

| 0.2 | 12/24 (0.50) [0.291, 0.709] | 1/24 (0.04) [0.001, 0.211] |
|---|---|---|
| 0.3 | 24/24 (1.00) [0.858, 1.000] | 1/24 (0.04) [0.001, 0.211] |
| 0.4 | 24/24 (1.00) [0.858, 1.000] | 2/24 (0.08) [0.010, 0.270] |
| 0.5 | 24/24 (1.00) [0.858, 1.000] | 0/24 (0.00) [0.000, 0.143] |
| 0.6 | 24/24 (1.00) [0.858, 1.000] | 1/24 (0.04) [0.001, 0.211] |
| 0.7 | 24/24 (1.00) [0.858, 1.000] | 12/24 (0.50) [0.291, 0.709] |
| 0.8 | 24/24 (1.00) [0.858, 1.000] | 21/24 (0.88) [0.676, 0.973] |
| 1.0 | 24/24 (1.00) [0.858, 1.000] | 24/24 (1.00) [0.858, 1.000] |

This design could not reject at f = 0 by construction and its MDEs are optimistic; it is kept only as a record (Table S6).

**Table S3d. Full identity-noise grid.**

Robustness of the battery non-separation result and of the degree-preserving continuity check to simulated buyer identity noise. Three perturbation modes are applied to buyer identities at six noise levels ε (plus the noiseless baseline ε = 0): "merge" collapses pairs of distinct buyers onto one identity, as exact homonyms would (their rate is unmeasured, so merge rates are simulated), "split" divides one buyer's rows between two identities, as near-duplicate spellings would (measured near-duplicate rate 0.7%, the grid value ε = 0.007), and "mixed" applies merging and splitting together (ε/2 of each). Twenty independent draws are generated per mode per ε (one draw at ε = 0, the exact reproduction of the canonical graphs). For each draw, the table reports whether the five-metric Holm battery separates 2019 from 2020, whether Portrait Divergence rejects, and whether the degree-preserving continuity check is significant, together with 95% Clopper–Pearson confidence intervals on each proportion.

| Perturbation mode | ε | Draws | Battery separations (n/draws) | Battery 95% CI | PD rejections (n/draws) | PD 95% CI | Continuity evaluable (n/draws) | Continuity significant (n/evaluable) | Continuity 95% CI | Median continuity z |
|---|---|---|---|---|---|---|---|---|---|---|
| merge | 0 | 1 | 0/1 | [0.000, 0.975] | 0/1 | [0.000, 0.975] | 1/1 | 1/1 | [0.025, 1.000] | 2.217 |
| merge | 0.007 | 20 | 0/20 | [0.000, 0.168] | 1/20 | [0.001, 0.249] | 20/20 | 4/20 | [0.057, 0.437] | 0.579 |
| merge | 0.010 | 20 | 1/20 | [0.001, 0.249] | 1/20 | [0.001, 0.249] | 20/20 | 2/20 | [0.012, 0.317] | −0.591 |
| merge | 0.020 | 20 | 0/20 | [0.000, 0.168] | 0/20 | [0.000, 0.168] | 20/20 | 1/20 | [0.001, 0.249] | −0.196 |

| merge | 0.050 | 20 | 2/20 | [0.012, 0.317] | 4/20 | [0.057, 0.437] | 20/20 | 0/20 | [0.000, 0.168] | −0.282 |
|---|---|---|---|---|---|---|---|---|---|---|
| merge | 0.100 | 20 | 1/20 | [0.001, 0.249] | 4/20 | [0.057, 0.437] | 20/20 | 1/20 | [0.001, 0.249] | −0.571 |
| merge | 0.200 | 20 | 1/20 | [0.001, 0.249] | 2/20 | [0.012, 0.317] | 15/20 | 1/15 | [0.002, 0.319] | −0.459 |
| split | 0 | 1 | 0/1 | [0.000, 0.975] | 0/1 | [0.000, 0.975] | 1/1 | 1/1 | [0.025, 1.000] | 2.297 |
| split | 0.007 | 20 | 0/20 | [0.000, 0.168] | 0/20 | [0.000, 0.168] | 20/20 | 19/20 | [0.751, 0.999] | 2.256 |
| split | 0.010 | 20 | 0/20 | [0.000, 0.168] | 0/20 | [0.000, 0.168] | 20/20 | 17/20 | [0.621, 0.968] | 2.226 |
| split | 0.020 | 20 | 0/20 | [0.000, 0.168] | 0/20 | [0.000, 0.168] | 20/20 | 14/20 | [0.457, 0.881] | 2.192 |
| split | 0.050 | 20 | 0/20 | [0.000, 0.168] | 1/20 | [0.001, 0.249] | 20/20 | 11/20 | [0.315, 0.769] | 2.063 |
| split | 0.100 | 20 | 0/20 | [0.000, 0.168] | 1/20 | [0.001, 0.249] | 20/20 | 7/20 | [0.154, 0.592] | 1.435 |
| split | 0.200 | 20 | 0/20 | [0.000, 0.168] | 1/20 | [0.001, 0.249] | 20/20 | 5/20 | [0.087, 0.491] | 0.558 |
| mixed | 0 | 1 | 0/1 | [0.000, 0.975] | 0/1 | [0.000, 0.975] | 1/1 | 1/1 | [0.025, 1.000] | 2.309 |
| mixed | 0.007 | 20 | 0/20 | [0.000, 0.168] | 0/20 | [0.000, 0.168] | 20/20 | 8/20 | [0.191, 0.639] | 1.442 |
| mixed | 0.010 | 20 | 0/20 | [0.000, 0.168] | 0/20 | [0.000, 0.168] | 20/20 | 7/20 | [0.154, 0.592] | 1.119 |
| mixed | 0.020 | 20 | 1/20 | [0.001, 0.249] | 0/20 | [0.000, 0.168] | 20/20 | 2/20 | [0.012, 0.317] | 0.479 |
| mixed | 0.050 | 20 | 0/20 | [0.000, 0.168] | 3/20 | [0.032, | 20/20 | 1/20 | [0.001, 0.249] | −0.443 |
| | | | | | | 0.379] | | | | |
| mixed | 0.100 | 20 | 3/20 | [0.032, 0.379] | 1/20 | [0.001, 0.249] | 20/20 | 3/20 | [0.032, 0.379] | −0.354 |
| mixed | 0.200 | 20 | 1/20 | [0.001, 0.249] | 0/20 | [0.000, 0.168] | 20/20 | 2/20 | [0.012, 0.317] | 0.254 |

*Note: ε = 0 rows are the noiseless baseline (1 exact reproduction of the canonical graphs), reported for reference and not part of the 20-draw noise grid. At ε = 0.20 under merging, 5 of 20 draws could not be rewired within the swap budget; their continuity tests are reported as unevaluable, and the continuity columns for that row are counted out of 15 evaluable draws.*

## Table S3e. Planted-effect power under further partitions of the anchor brands (post hoc; 14,235 and 10,399 buyer-year baskets per replicate; redirection planted in the second group; 100 replicates per f and partition; inner relabel null R = 999).

| Partition | Block sizes | f | Replicates | Battery rejections (power) [95% CI] | Battery defined share | PD rejections (power) [95% CI] | Density | Mean degree | Clustering | Modularity | Assortativity |
|---|---|---|---|---|---|---|---|---|---|---|---|
| 0 (Design P) | 30/30 | 0.2 | 100 | 33/100 (0.33) [0.239, 0.431] | 1.00 | 7/100 (0.07) [0.029, 0.139] | 0.11 | 0.11 | 0.14 | 0.02 | 0.19 |
| 0 (Design P) | 30/30 | 0.3 | 100 | 83/100 (0.83) [0.742, 0.898] | 1.00 | 17/100 (0.17) [0.102, 0.258] | 0.31 | 0.30 | 0.63 | 0.10 | 0.74 |
| 0 (Design P) | 30/30 | 0.4 | 100 | 97/100 (0.97) [0.915, 0.994] | 1.00 | 26/100 (0.26) [0.177, 0.357] | 0.71 | 0.58 | 0.89 | 0.50 | 0.96 |
| 1 | 30/30 | 0.2 | 100 | 29/100 (0.29) [0.204, 0.389] | 1.00 | 6/100 (0.06) [0.022, 0.126] | 0.09 | 0.09 | 0.15 | 0.04 | 0.17 |
| 1 | 30/30 | 0.3 | 100 | 89/100 (0.89) [0.812, 0.944] | 1.00 | 24/100 (0.24) [0.160, 0.336] | 0.34 | 0.29 | 0.62 | 0.15 | 0.79 |
| 1 | 30/30 | 0.4 | 100 | 99/100 (0.99) [0.946, 1.000] | 1.00 | 27/100 (0.27) [0.186, 0.368] | 0.72 | 0.60 | 0.92 | 0.50 | 0.97 |
| 2 | 30/30 | 0.2 | 100 | 28/100 (0.28) [0.195, 0.379] | 1.00 | 8/100 (0.08) [0.035, 0.152] | 0.09 | 0.05 | 0.13 | 0.04 | 0.19 |
| 2 | 30/30 | 0.3 | 100 | 85/100 (0.85) [0.765, 0.914] | 1.00 | 20/100 (0.20) [0.127, 0.292] | 0.30 | 0.29 | 0.61 | 0.09 | 0.78 |
| 2 | 30/30 | 0.4 | 100 | 98/100 (0.98) [0.930, 0.998] | 1.00 | 25/100 (0.25) [0.169, 0.347] | 0.71 | 0.58 | 0.91 | 0.46 | 0.96 |
| 3 | 30/30 | 0.2 | 100 | 31/100 (0.31) [0.221, 0.410] | 1.00 | 9/100 (0.09) [0.042, 0.164] | 0.12 | 0.08 | 0.16 | 0.03 | 0.20 |
| 3 | 30/30 | 0.3 | 100 | 85/100 (0.85) [0.765, 0.914] | 1.00 | 20/100 (0.20) [0.127, 0.292] | 0.26 | 0.24 | 0.59 | 0.13 | 0.73 |
| 3 | 30/30 | 0.4 | 100 | 97/100 (0.97) [0.915, 0.994] | 1.00 | 32/100 (0.32) [0.230, 0.421] | 0.67 | 0.50 | 0.92 | 0.50 | 0.96 |
| 4 | 30/30 | 0.2 | 100 | 38/100 (0.38) [0.285, 0.483] | 1.00 | 9/100 (0.09) [0.042, 0.164] | 0.08 | 0.07 | 0.17 | 0.02 | 0.24 |
| 4 | 30/30 | 0.3 | 100 | 84/100 (0.84) [0.753, 0.906] | 1.00 | 24/100 (0.24) [0.160, 0.336] | 0.28 | 0.30 | 0.56 | 0.12 | 0.75 |
| 4 | 30/30 | 0.4 | 100 | 100/100 (1.00) [0.964, 1.000] | 1.00 | 30/100 (0.30) [0.212, 0.400] | 0.67 | 0.56 | 0.91 | 0.50 | 1.00 |
| 5 | 30/30 | 0.2 | 100 | 36/100 (0.36) [0.266, 0.462] | 1.00 | 8/100 (0.08) [0.035, 0.152] | 0.10 | 0.07 | 0.15 | 0.04 | 0.26 |
| 5 | 30/30 | 0.3 | 100 | 85/100 (0.85) [0.765, 0.914] | 1.00 | 18/100 (0.18) [0.110, 0.269] | 0.21 | 0.22 | 0.51 | 0.13 | 0.79 |
| 5 | 30/3 | 0. | 100 | 98/100 | 1.00 | 33/100 | 0.71 | 0.64 | 0.91 | 0.48 | 0.95 |
| | 0 | 4 | | (0.98) [0.930, 0.998] | | (0.33) [0.239, 0.431] | | | | | |
| 6 | 30/30 | 0.2 | 100 | 31/100 (0.31) [0.221, 0.410] | 1.00 | 7/100 (0.07) [0.029, 0.139] | 0.09 | 0.09 | 0.17 | 0.03 | 0.23 |
| 6 | 30/30 | 0.3 | 100 | 92/100 (0.92) [0.848, 0.965] | 1.00 | 18/100 (0.18) [0.110, 0.269] | 0.31 | 0.27 | 0.65 | 0.16 | 0.82 |
| 6 | 30/30 | 0.4 | 100 | 98/100 (0.98) [0.930, 0.998] | 1.00 | 30/100 (0.30) [0.212, 0.400] | 0.76 | 0.63 | 0.93 | 0.53 | 0.96 |
| 7 | 30/30 | 0.2 | 100 | 35/100 (0.35) [0.257, 0.452] | 1.00 | 10/100 (0.10) [0.049, 0.176] | 0.08 | 0.08 | 0.20 | 0.04 | 0.23 |
| 7 | 30/30 | 0.3 | 100 | 88/100 (0.88) [0.800, 0.936] | 1.00 | 23/100 (0.23) [0.152, 0.325] | 0.28 | 0.27 | 0.64 | 0.13 | 0.78 |
| 7 | 30/30 | 0.4 | 100 | 97/100 (0.97) [0.915, 0.994] | 1.00 | 24/100 (0.24) [0.160, 0.336] | 0.64 | 0.58 | 0.92 | 0.48 | 0.97 |
| 8 | 30/30 | 0.2 | 100 | 32/100 (0.32) [0.230, 0.421] | 1.00 | 6/100 (0.06) [0.022, 0.126] | 0.10 | 0.08 | 0.15 | 0.05 | 0.21 |
| 8 | 30/30 | 0.3 | 100 | 87/100 (0.87) [0.788, 0.929] | 1.00 | 19/100 (0.19) [0.118, 0.281] | 0.32 | 0.29 | 0.66 | 0.11 | 0.80 |
| 8 | 30/30 | 0.4 | 100 | 97/100 (0.97) [0.915, 0.994] | 1.00 | 22/100 (0.22) [0.143, 0.314] | 0.68 | 0.56 | 0.92 | 0.45 | 0.96 |
| 9 | 30/30 | 0.2 | 100 | 33/100 (0.33) [0.239, 0.431] | 1.00 | 8/100 (0.08) [0.035, 0.152] | 0.08 | 0.06 | 0.16 | 0.03 | 0.25 |
| 9 | 30/30 | 0.3 | 100 | 83/100 (0.83) | 1.00 | 20/100 (0.20) | 0.33 | 0.31 | 0.61 | 0.17 | 0.77 |
| | | | | [0.742, 0.898] | | [0.127, 0.292] | | | | | |
| 9 | 30/30 | 0.4 | 100 | 98/100 (0.98) [0.930, 0.998] | 1.00 | 29/100 (0.29) [0.204, 0.389] | 0.71 | 0.57 | 0.87 | 0.49 | 0.98 |
| 10 | 30/30 | 0.2 | 100 | 34/100 (0.34) [0.248, 0.442] | 1.00 | 6/100 (0.06) [0.022, 0.126] | 0.07 | 0.07 | 0.14 | 0.02 | 0.27 |
| 10 | 30/30 | 0.3 | 100 | 86/100 (0.86) [0.776, 0.921] | 1.00 | 23/100 (0.23) [0.152, 0.325] | 0.27 | 0.24 | 0.57 | 0.14 | 0.75 |
| 10 | 30/30 | 0.4 | 100 | 98/100 (0.98) [0.930, 0.998] | 1.00 | 29/100 (0.29) [0.204, 0.389] | 0.70 | 0.57 | 0.91 | 0.48 | 0.97 |
| category | 26/34 | 0.2 | 100 | 51/100 (0.51) [0.408, 0.611] | 1.00 | 8/100 (0.08) [0.035, 0.152] | 0.11 | 0.08 | 0.26 | 0.07 | 0.37 |
| category | 26/34 | 0.3 | 100 | 90/100 (0.90) [0.824, 0.951] | 1.00 | 24/100 (0.24) [0.160, 0.336] | 0.35 | 0.28 | 0.68 | 0.28 | 0.83 |
| category | 26/34 | 0.4 | 100 | 99/100 (0.99) [0.946, 1.000] | 1.00 | 31/100 (0.31) [0.221, 0.410] | 0.82 | 0.66 | 0.89 | 0.71 | 0.98 |

Partition 0 is the partition of Table S3a. Partitions 1–10 split the same 60 anchor brands at random into two blocks of 30 with further seeds; the category partition puts the 26 anchors whose modal product category is watch or accessory in one block and the other 34 in the other, and a mover chooses among the blocks large enough for its breadth (no unit held more than 11 anchor brands). The cell seeds are those of Table S3a, so the random partitions share their groups, movers and inner draws and differ only in where movers are redirected; their spread is therefore not an interval over partitions. At $f = 0.3$ power was at least 0.8 under all 11 random partitions (range 0.83–0.92, median 0.85) and 0.90 under the category partition; for every partition the first tested $f$ with battery power at least 0.8 and defined share at least 0.95 is 0.3. The per-metric columns are Holm rejection rates. $f = 0$ does not depend on the partition (Table S3a).

**Table S3f. Observed 2019-to-2020 differences beside the mean differences produced by a planted redirection (descriptive; no test).**

| Source | f | Density | Mean degree | Clustering | Modularity | Assortativity |
|---|---|---|---|---|---|---|
| observed 2020 minus 2019 | – | −0.017 | −1.76 | −0.128 | +0.029 | −0.005 |
| Design P | 0 | −0.002 | −0.72 | −0.017 | +0.008 | +0.007 |
| Design P | 0.1 | −0.009 | −0.98 | −0.063 | +0.006 | +0.030 |
| Design P | 0.2 | −0.035 | −1.45 | −0.135 | +0.017 | +0.074 |
| Design P | 0.3 | −0.059 | −1.90 | −0.233 | +0.040 | +0.175 |
| Design P | 0.4 | −0.080 | −2.25 | −0.327 | +0.085 | +0.288 |
| mean over the 11 random partitions | 0.2 | −0.033 | −1.42 | −0.132 | +0.017 | +0.079 |
| category partition | 0.2 | −0.043 | −1.49 | −0.153 | +0.033 | +0.112 |
| mean over the 11 random partitions | 0.3 | −0.058 | −1.90 | −0.232 | +0.041 | +0.183 |
| category partition | 0.3 | −0.062 | −1.84 | −0.239 | +0.069 | +0.217 |
| mean over the 11 random partitions | 0.4 | −0.079 | −2.24 | −0.326 | +0.086 | +0.299 |
| category partition | 0.4 | −0.083 | −2.17 | −0.319 | +0.126 | +0.325 |

Design P rows are means over the 100 replicates of Table S3a; the other rows average the 11 random partitions (Design P included) or give the category partition (Table S3e). The f = 0 row is the no-planting baseline: the nulls expect a fall in mean degree because the second group is smaller. Differences are 2020 group minus 2019 group; displayed values are rounded once.

## Table S4. Public-benchmark construction, size check and diagnostics

The UCI Online Retail II dataset [45], released under CC BY 4.0 at https://doi.org/10.24432/C5CG6D, is used to transfer the inference protocol to a public benchmark under an adapted edge threshold, since the literal shared-buyer threshold used on the proprietary data ($\tau$ = 2, 3 or 5) produces a complete graph on this denser retail dataset. Table S4a shows the pre-specified five-step cleaning ledger from the 1,067,371 raw transaction rows to the 775,543 rows used in the analysis. Table S4b shows the construction grid searched to select the adapted threshold $\tau^*$, together with the three candidate construction rules and the pre-specified selection rule. Table S4c shows the implementation-size check: the rate at which each test rejects on same-window half-splits of the data, which a correct implementation keeps at or below the nominal level; a rate below it means the test does not over-reject, and a conservative size also lowers power. Table S4e shows the planted-effect power curve at split group sizes, which replaces the superseded fixed-base curve shown beneath it. Table S4d shows three diagnostics: an intended UK-versus-non-UK positive control, and two versions (relabel and paired-swap) of a W1-versus-W2 comparison across the two one-year windows of the benchmark, which is not a shock comparison and is reported only to check that the paired-swap null again gives a smaller Portrait Divergence p-value than the relabel null, as it does on the proprietary data.

**Table S4a. UCI Online Retail II pre-specified cleaning ledger.**

| Step | Rows in | Rows out | Rows removed | Leading removed codes (count) |
|---|---|---|---|---|
| raw | 1,067,371 | 1,067,371 | 0 | n/a |
| 1 invoice starts C/A | 1,067,371 | 1,047,871 | 19,500 | n/a |
| 2 missing Customer ID | 1,047,871 | 805,620 | 242,251 | n/a |
| 3 Quantity<=0 or Price<=0 | 805,620 | 805,549 | 71 | n/a |
| 4 StockCode pattern | 805,549 | 801,559 | 3,990 | POST:1838; 15056BL:838; M:709; C2:253; 79323LP:159; 79323GR:76; BANK CHARGES:32; ADJUST:32; PADS:17; DOT:16; TEST001:9; D:5; ADJUST2:3; SP1002:2; TEST002:1 |
| 5 exact duplicates | 801,559 | 775,543 | 26,016 | n/a |

**Table S4b. Adapted-threshold construction grid (candidate rule × τ, both windows).**

| Candidate rule | τ | Window | Customers | Nodes | Edges | Density | Rank-60 tie |
|---|---|---|---|---|---|---|---|
| literal threshold | 3 | W1 | 4,239 | 60 | 1,770 | 1.000 | no |
| literal threshold | 3 | W2 | 4,293 | 60 | 1,770 | 1.000 | no |
| literal threshold | 2 | W1 | 4,239 | 60 | 1,770 | 1.000 | no |
| literal threshold | 2 | W2 | 4,293 | 60 | 1,770 | 1.000 | no |
| literal threshold | 5 | W1 | 4,239 | 60 | 1,770 | 1.000 | no |

| literal threshold | 5 | W2 | 4,293 | 60 | 1,770 | 1.000 | no |
|---|---|---|---|---|---|---|---|
| adapted threshold, all customers | 63 | W1 | 4,239 | 60 | 1,679 | 0.949 | no |
| adapted threshold, all customers | 63 | W2 | 4,293 | 60 | 1,558 | 0.880 | no |
| adapted threshold, all customers | 93 | W1 | 4,239 | 60 | 1,042 | 0.589 | no |
| adapted threshold, all customers | 93 | W2 | 4,293 | 60 | 904 | 0.511 | no |
| adapted threshold, all customers | 156 | W1 | 4,239 | 59 | 221 | 0.129 | no |
| adapted threshold, all customers | 156 | W2 | 4,293 | 52 | 211 | 0.159 | no |
| adapted threshold, retail tier | 6 | W1 | 2,124 | 60 | 1,328 | 0.750 | no |
| adapted threshold, retail tier | 6 | W2 | 2,147 | 60 | 1,318 | 0.745 | yes |
| adapted threshold, retail tier | 8 | W1 | 2,124 | 60 | 1,009 | 0.570 | no |
| adapted threshold, retail tier | 8 | W2 | 2,147 | 60 | 1,002 | 0.566 | yes |
| adapted threshold, retail tier | 14 | W1 | 2,124 | 60 | 399 | 0.225 | no |
| adapted threshold, retail tier | 14 | W2 | 2,147 | 60 | 413 | 0.233 | yes |

Candidate construction rules, evaluated in this order: literal threshold (the proprietary rule, $\tau = 3$, then 2, then 5); adapted threshold on all customers ($\tau$ = *the smallest* $\tau \geq 3$ *with W1 density* $\leq 0.60$*, reported with* $\lfloor 0.67\tau \rceil$ and $\lceil 1.67\tau \rceil$*); adapted threshold on the retail tier (customers whose window total quantity is at or below that window's median,* $\tau$ derived by the same rule). A rule is feasible when both windows give at least 25 nodes and density between 0.05 and 0.60, and the first feasible rule is the analysis construction (pre-specified). The literal thresholds all give complete graphs (density 1.000), so the adapted threshold on all customers carries the analysis with $\tau^* = 93$ (W1 density 0.589, W2 density 0.511).

**Table S4c. Implementation-size check: same-window half-split rejection rates (Clopper–Pearson 95% CI).**

| Window | Test | Splits | Defined | Rejections | Rate | 95% CI |
|---|---|---|---|---|---|---|
| W1 | battery | 500 | 500 | 13 | 0.026 | [0.014, 0.044] |
| W1 | PD | 500 | 500 | 19 | 0.038 | [0.023, 0.059] |
| W2 | battery | 500 | 500 | 14 | 0.028 | [0.015, 0.047] |
| W2 | PD | 500 | 500 | 24 | 0.048 | [0.031, 0.071] |
| pooled | battery | 1,000 | 1,000 | 27 | 0.027 | [0.018, 0.039] |
| pooled | PD | 1,000 | 1,000 | 43 | 0.043 | [0.031, 0.057] |

**Table S4d. Benchmark diagnostics.**

| Diagnostic | Defined | Observed PD | PD p | PD rejects | Min Holm p (five-metric battery) | Battery rejects |
|---|---|---|---|---|---|---|
| UK vs non-UK customers (W2) and its median-total-quantity proxy | not evaluable | n/a | n/a | n/a | n/a | n/a |
| W1 vs W2, relabel null | yes | 0.433 | 0.092 | no | 0.670 | no |
| W1 vs W2, paired-swap null | yes | 0.433 | 0.033 | yes | 0.255 | no |

The UK versus non-UK positive control is not evaluable: the non-UK graph and its fallback proxy (customers above versus at or below the median total quantity) both fall below 20 nodes at the adapted threshold, so no permutation test could be constructed for it.

*Table S4e. Benchmark planted-effect power at split group sizes (8,532 W1 and W2 customer-window units split into 4,239 and 4,293; redirection planted in the second group; $\tau = 93$; 100 replicates per f; inner relabel null R = 199).**

| f | Replicates | Battery rejections (power) [95% CI] | Battery defined share | PD rejections (power) [95% CI] | Density | Mean degree | Clustering | Modularity | Assortativity |
|---|---|---|---|---|---|---|---|---|---|
| 0 | 100 | 2/100 (0.02) [0.002, | 1.00 | 4/100 (0.04) [0.011, | 0.00 | 0.00 | 0.00 | 0.01 | 0.01 |
| | | 0.070] | | 0.099] | | | | | |
| 0.05 | 100 | 1/100 (0.01) [0.000, 0.054] | 1.00 | 5/100 (0.05) [0.016, 0.113] | 0.00 | 0.00 | 0.00 | 0.01 | 0.00 |
| 0.1 | 100 | 8/100 (0.08) [0.035, 0.152] | 1.00 | 5/100 (0.05) [0.016, 0.113] | 0.00 | 0.00 | 0.05 | 0.03 | 0.02 |
| 0.2 | 100 | 80/100 (0.80) [0.708, 0.873] | 1.00 | 19/100 (0.19) [0.118, 0.281] | 0.04 | 0.04 | 0.18 | 0.01 | 0.77 |
| 0.3 | 100 | 100/100 (1.00) [0.964, 1.000] | 1.00 | 78/100 (0.78) [0.686, 0.857] | 0.10 | 0.10 | 0.01 | 0.58 | 1.00 |
| 0.45 | 100 | 100/100 (1.00) [0.964, 1.000] | 1.00 | 100/100 (1.00) [0.964, 1.000] | 0.23 | 0.23 | 1.00 | 1.00 | 1.00 |
| 0.6 | 100 | 100/100 (1.00) [0.964, 1.000] | 1.00 | 100/100 (1.00) [0.964, 1.000] | 0.40 | 0.40 | 1.00 | 1.00 | 1.00 |
| 0.8 | 100 | 98/100 (0.98) [0.930, 0.998] | 0.98 | 100/100 (1.00) [0.964, 1.000] | 0.65 | 0.65 | 0.98 | 0.98 | 0.98 |
| 1.0 | 100 | 0/100 (0.00) [0.000, 0.036] | 0.00 | 100/100 (1.00) [0.964, 1.000] | 0.00 | 0.00 | 0.00 | 0.00 | 0.00 |

Battery MDE f = 0.2; PD MDE f = 0.45; the pre-specified two-part criterion (battery MDE < 1 and battery MDE ≤ PD MDE) is met. Undefined battery results (degree assortativity not computable) count as non-rejections. Per replicate, on average 3,183 of the 4,293 second-group units (74.2%) were eligible (2 to 30 of the 60 anchor products) and 48 held more than 30 anchor products and were not redirected; on the proprietary data 593 of 10,399 second-group baskets (5.7%) were eligible (Table S3a). f is a share of eligible units, so it is not comparable across the two datasets.

Superseded (fixed-base planted contrast; no between-sample variability): the pre-specified A1-2 curve (100 replicates per f).

| **f** | **Battery power** | **PD power** | **Defined share** |
|---|---|---|---|

| 0 | 0.00 | 0.00 | 1.00 |
|---|---|---|---|
| 0.05 | 0.00 | 0.02 | 1.00 |
| 0.1 | 0.00 | 0.04 | 1.00 |
| 0.2 | 1.00 | 0.21 | 1.00 |
| 0.3 | 1.00 | 0.99 | 1.00 |
| 0.45 | 1.00 | 1.00 | 1.00 |
| 0.6 | 1.00 | 1.00 | 1.00 |
| 0.8 | 0.01 | 0.01 | 0.01 |
| 1.0 | 0.00 | 0.00 | 0.00 |

## Table S5. Size-matched removal and D′ by object

Per-object results of the size-matched buyer-removal comparison (a pre-specified gate) and its breadth-matched diagnostic counterpart (D′). For each object (the pooled 2018–2021 graph and each individual year), all cross-border buyers are removed from the full buyer-brand network and the fraction of backbone edges retained (ret_X) is compared against 500 draws that instead remove an equal number of randomly selected domestic buyers (ret_D) or, for D′, domestic buyers matched to the cross-border group's purchase-breadth distribution (ret_D′). Δ = ret_X − ret_D and Δ′ = ret_X − ret_D′ are reported with their means, 2.5–97.5 percentile ranges over the draws, and the count of draws exceeding zero. The gate was pre-specified with the hypothesis that Δ would be positive (cross-border buyers contributing less per head); every object shows the opposite sign, so the gate is reported as failed rather than passed. D′ is a diagnostic and is not available for 2019.

| Object | n cross-border | n domestic | Cross-border retention (ret_X) | Domestic-draw retention mean [2.5–97.5%] | Δ = ret_X − ret_D mean [2.5–97.5%] | Draws with Δ > 0 (of 500) | D′ (breadth-matched) retention mean [2.5–97.5%] | Δ′ mean [2.5–97.5%] | Draws with Δ′ > 0 (of 500) |
|---|---|---|---|---|---|---|---|---|---|
| pooled | 11,367 | 31,595 | 0.755 | 0.827 [0.767, 0.879] | −0.071 [−0.124, −0.011] | 6/500 | 0.692 [0.634, 0.749] | 0.063 [0.006, 0.121] | 492/500 |
| 2018 | 3,458 | 9,344 | 0.662 | 0.793 [0.669, 0.900] | −0.132 [−0.238, −0.008] | 9/500 | 0.521 [0.462, 0.623] | 0.140 [0.038, 0.200] | 497/500 |
| 2019 | 5,449 | 8,786 | 0.584 | 0.758 [0.686, 0.839] | −0.174 [−0.255, −0.102] | 0/500 | not available | not available | not available |
| 2020 | 2,016 | 8,383 | 0.750 | 0.886 [0.810, 0.952] | −0.136 [−0.202, −0.060] | 0/500 | 0.717 [0.643, 0.810] | 0.033 [−0.060, 0.107] | 363/500 |

| 2021 | 1,265 | 7,955 | 0.722 | 0.905 [0.785, 0.972] | −0.183 [−0.250, −0.063] | 0/500 | 0.684 [0.573, 0.806] | 0.039 [−0.083, 0.149] | 351/500 |
|---|---|---|---|---|---|---|---|---|---|

*Note: D′ (breadth-matched domestic draws) is not available for 2019 (one domestic breadth stratum was smaller than its cross-border counterpart). The size-matched removal is a pre-specified gate that failed (sign opposite to the pre-specified hypothesis in every object); D′ is a diagnostic. Differences are computed at full precision and rounded once, so a displayed Δ or Δ′ can differ by 0.001 from the difference of the displayed retentions.*

## Table S6. Analysis record

Every analysis reported in the paper, with its status, the date its specification was fixed, its decision criterion, its outcome and any deviation from the written specification. Status: pre-specified = specified in a dated internal written protocol frozen before the analysis ran (no protocol was publicly registered); analysis plan after an initial comparison = the primary construction and battery, written in the analysis plan after an initial comparison at this construction had found no separation, with the reported values from later reruns; fixed before the reported run = design fixed before the reported run, without a separate protocol; diagnostic = reported without a decision criterion (secondary: supporting evidence only; fixed before the run: design fixed before it ran); descriptive = summary statistic without a test; planned, not run = specified and deliberately not run, for the reason given; superseded = replaced by a later analysis and kept only as a record; corrective analysis, fixed before its results = specified after a defect in an earlier design was found and before any of its own results were seen; implementation correction, after the earlier results were seen = a code correction made after earlier results were seen, with every affected analysis rerun; post hoc (used in the Deviation column) = changed after results were seen; post hoc sensitivity analysis = specified after the primary results were seen and fixed in a written protocol before it ran. Dated protocol copies are available to editors and reviewers.

| Analysis | Status | Date fixed | Criterion | Outcome | Deviation |
|---|---|---|---|---|---|
| Primary comparison, 2019 vs 2020, relabel null (Table 1) | Analysis plan after an initial comparison | Analysis plan 2026-07-03 (two-sided difference tests, Holm primary, Benjamini–Hochberg secondary, R = 2,000); five-metric family restated in the protocols of 2026-09-23 and 2026-09-24 | The years separate if any Holm-adjusted two-sided p < 0.05 | No metric separates (smallest Holm p 0.407); PD one-sided p = 0.259 | Six → five metrics (post hoc amendment): the plan named six metrics; after the six-metric analysis the sixth, the largest-connected-component fraction, which was 1.000 in both observed graphs and in all 2,000 null draws (raw p |
| | | | | | = 1.000), was removed. Under the original six-metric family the smallest Holm p is 0.489 (relabel) and 0.117 (paired swap), so neither family separates the years. The 2026-09-24 run reproduced the earlier run's differences and raw p-values metric for metric; the reported values are the 2026-09-27 rerun with one graph builder, one tie rule and merged duplicate brand spellings (rows Node labels, Tie rule and Brand spellings) |
| Paired-swap null | Fixed before the reported run | 2026-09-24 | As for the primary comparison | No metric separates (smallest Holm p 0.097); PD one-sided p = 0.118 | Rerun 2026-09-26 with one graph builder and one tie rule (earlier: smallest Holm p 0.092, PD p = 0.120) and 2026-09-27 with merged brand spellings (smallest Holm p 0.095 → 0.097, PD p the same). Replaces an earlier buyer-clustered null whose second group held 9,461 instead of 10,399 buyers (PD p = 0.418 in that version) |
| Calibration of the paired-swap null on the public benchmark | Planned, not run | Drafted 2026-09-24 | Rejection rates of battery and PD ≤ 0.07 (size held) or either ≥ 0.09 (anti-conservative) | Not run | Its pseudo-period construction makes the baskets exchangeable by design, so it would test only the implementation; the null's exchangeability assumption is stated in Methods instead. The size of the whole paired-swap procedure was measured on the proprietary data in the paired-swap check (4/100 battery, 5/100 PD at f = 0; Table S3b) |
| Union brand set, relabel null | Fixed before the reported run | 2026-09-24 | Any Holm $p < 0.05$ qualifies the non-detection | No metric separates (Holm p 0.490–1.000); PD $p = 0.109$ | Replaces a version whose observed side used the per-year top-60 graphs; the 2026-09-26 single-builder rerun changed only the modularity raw p (0.487 → 0.478), and the 2026-09-27 brand-spelling rerun changed it to 0.480 |
| Resolution sweep (four constructions × two nulls) | Pre-specified | 2026-09-24 | Parity with the primary run; if any construction separates under either null, name it and restrict the null claim to the primary construction | Rerun 2026-09-26 under one tie rule and 2026-09-27 with merged brand spellings: the battery separates nowhere under either null or either tie rule (smallest Holm p 0.097); at top-111, ≥ 1 PD $p = 0.0445$ (relabel) | The earliest result (PD $p = 0.042$ at top-80, ≥ 2, paired swap; 0.091 relabel) came from observed and null graphs ranked by different sort routines and is superseded (rows Node labels and Tie rule). Before |

| | | | | and 0.0225 (paired swap), 0.0460 and 0.0245 under the partial-sort rule, below 0.05 under both rules but nominal (Bonferroni 0.180 over eight tests, 0.090 over the four paired-swap tests); at top-80, ≥ 2 the paired-swap PD p is 0.173 against 0.0430 under the partial sort and is reported as tie-dependent; the reporting rule was applied and the PD claim is restricted to the primary construction | the brand-spelling merge the smallest PD p was 0.0465 (top-111, ≥ 1, paired swap; 0.0555 under the partial sort), and top-80, ≥ 2 gave 0.177 (paired swap) and 0.281 (relabel) under the brand-name rule (row Brand spellings) |
|---|---|---|---|---|---|
| Threshold sweep (t = 2, 3, 5) | Diagnostic | Rerun 2026-09-24 | None | The battery separates at no threshold | None |
| Identity-noise grid | Pre-specified | 2026-09-24 | A battery sensitivity claim needs ≥ 16 of 20 separations at ε = 0.007 or 0.01 in one mode (10–15 unresolved); continuity is reported by mode | Battery 0/20 at ε = 0.007 in every mode; continuity significant in 4/20 (merge), 8/20 (mixed) and 19/20 (split) draws at ε = 0.007. The 2026-09-26 single-builder rerun changed one cell (mixed, ε = 0.05: battery 1/20 → 0/20); the 2026-09-27 brand-spelling rerun changed no count | Note written 2026-09-24, after the first rewiring failure and before any affected draw was rerun: a draw whose rewiring exhausts the swap budget is unevaluable for continuity (5 draws, merge, ε = 0.20) |
| Positive control on the exposure cells | Diagnostic, fixed before the run | 2026-09-24 | Report battery and PD rejections out of 6; no battery | Battery 6/6 (smallest Holm p per pair 0.0025– | Replaces a version that ranked brands by buyer count |
| | | | rejection would be reported as a failed check | 0.0325); PD 2/6 | instead of summed sale value; the 2026-09-26 single-builder rerun left the counts as they were and changed the smallest Holm p range (earlier 0.0025–0.0475); after the 2026-09-27 brand-spelling merge the Off-Local cell has 45 nodes (earlier 44) and the counts and range are the same |
| Planted-effect power curve, proprietary data (fixed-base design) | Superseded | Five-metric recomputation fixed 2026-09-23 | MDE = smallest grid f with power ≥ 0.8 | Battery MDE 0.3, PD MDE 0.8 under the fixed-base design (Table S3c); superseded by the corrected design below | Six → five metrics, recomputed from the same draws as the six-metric curve; superseded 2026-09-26 (no between-sample variability) |
| Benchmark feasibility | Pre-specified | 2026-09-23 | First feasible candidate construction rule (Table S4b) | Passed: adapted threshold on all customers, τ* = 93 | None |
| Benchmark size check | Pre-specified | 2026-09-23 | All four rejection rates (battery and PD; all and defined splits) ≤ 0.07, and ≤ 5% of splits undefined | Passed: battery 0.027, PD 0.043; 1,000 of 1,000 splits defined | Rerun 2026-09-24 after three code fixes (power base ranked by row count; undefined results no longer scored as PD rejections; rank-60 ties logged); same verdict |
| Benchmark planted-effect power | Pre-specified | 2026-09-23 | Battery MDE < 1 and battery MDE ≤ PD MDE | Passed under the fixed-base design (battery MDE 0.2, PD MDE 0.3); superseded by the split-sample curve below (Table S4e) | Undefined results count as non-rejections, as the protocol specifies |

| Benchmark diagnostics (UK vs non-UK; W1 vs W2) | Diagnostic | 2026-09-23 | None | UK vs non-UK unevaluable; W1 vs W2: no battery separation, PD p = 0.092 (relabel) and 0.033 (paired swap) | The paired-swap variant was held until the paired-swap design was corrected (2026-09-24) and then run with it |
|---|---|---|---|---|---|
| Size-matched removal | Pre-specified | 2026-09-23 | Pooled, 2018 and 2019 objects each: mean Δ ≥ 0.10 and Δ > 0 in ≥ 450 of 500 draws | Failed, with the opposite sign (pooled Δ = −0.071; 6 of 500 draws above zero) | None; the 2026-09-24 rerun reproduced every draw, and the 2026-09-26 tie-rule rerun changed the pooled Δ in the third decimal (−0.072 → −0.071) with the same verdict |
| D′ (breadth-matched removal) | Diagnostic | 2026-09-23 | None | Pooled Δ′ = +0.063 (492 of 500 draws above zero); 2018 Δ′ = +0.140; unavailable for 2019 | Breadth strata redefined on the full-object top-60 before the 2026-09-24 rerun; the 2026-09-27 brand-spelling rerun changed the pooled Δ′ from +0.062 (494 of 500) |
| Modularity against the degree-preserving null | Diagnostic | Seed map 2026-09-24 | None | Every graph above its null (p = 0.002) | Canonical node order adopted after z was found to depend on node order (post hoc); z is reported with that caveat |
| Degree-preserving continuity | Diagnostic (secondary) | Rerun on canonical graphs 2026-09-24 | None | Forward z = 2.27, p = 0.0155 | The canonical rerun replaces a node-order-dependent version (z = 2.17) |
| Per-year subsampling bands | Descriptive | 2026-09-24 (finite-draw counts recorded) | None | Each Politis–Romano band contains its observed value by construction of the centring (20 of 20 | The *n*-out-of-*n* bootstrap was replaced by the Politis–Romano band (Table S2). Band scaling |

| | | | | metric-years); this is not evidence of coverage | corrected 2026-09-27 from √(m/n) to the without-replacement factor √(m/(n − m)) = 1, which widens every band |
|---|---|---|---|---|---|
| Planted-effect power at the real group sizes (Design P), composition-only alternative (Design S), season-matched sensitivity, paired-swap check | Corrective analysis, fixed before its results | 2026-09-26 | Size at f = 0: stop if either test rejects in ≥ 11 of 100 replicates; MDE = smallest grid f with power ≥ 0.8 and the test defined in ≥ 95% of replicates | Size at f = 0: battery 5/100, PD 2/100 (stop rule not triggered). Battery power 0.83 [0.742, 0.898] at f = 0.3 (0.86 [0.776, 0.921] before the 2026-09-27 brand-spelling rerun), MDE 0.3; PD MDE 1.0 (Table S3a). Composition-only model at a = 1: battery 15/100, PD 8/100 (below 0.80: the comparison has little power against a composition-only change). Season-matched: no battery separation (all Holm p 1.000), PD one-sided p = 0.019. Paired-swap check: battery 4/100 at f = 0 and 0.85 at f = 0.3 (Table S3b) | Replaces the fixed-base planted curve (a 2019 base against a planted copy of the same buyers: no between-sample variability, equal group sizes, no rejection possible at f = 0), which is kept in Table S3c as a superseded record; specified after that defect was found |
| Benchmark planted-effect power at split group sizes | Corrective analysis, fixed before its results | 2026-09-26 | As in the pre-specified rule: battery MDE < 1 and battery MDE ≤ PD MDE; size at f = 0 as above | Size at f = 0: battery 2/100, PD 4/100. Battery MDE 0.2 (power 0.80 [0.708, 0.873]), PD MDE 0.45; two-part criterion met (Table S4e) | Replaces the pre-specified fixed-base benchmark curve for the same reason; that curve's verdict stays on record |
| Node labels in the modularity statistic | Implementation correction, after the earlier results were seen | 2026-09-26 | Observed and null graphs from one builder | Modularity p changed in the third decimal (primary relabel test 0.4658 → 0.4648, paired swap 0.3773 → 0.3753, both measured before the tie-rule change); no Holm value changed at the primary construction | Observed graphs carried brand labels and null draws integer labels, and seeded Louvain depends on node order; every affected analysis was rerun |
| Tie rule at a rank cut-off | Implementation correction, after the earlier results were seen | 2026-09-26 | One rule everywhere: descending count, ties by ascending brand name | Resolution sweep rerun under this rule and, as a sensitivity analysis on the same permutation draws, under the partial-sort rule applied to observed and null graphs alike. After the 2026-09-27 brand-spelling merge, at top-80, ≥ 2 the rules keep different tied brands (2019: Casio and Montblanc under the brand-name rule, Mauboussin and Lucien Pellat-Finet under the partial sort; 2020: Chopard and Pomellato against Lucien Pellat-Finet and Red Valentino), giving 52 against 53 nodes in 2019 and paired-swap PD p = 0.173 against | Observed graphs and draws had used two different sort routines with no secondary key; at top-80, ≥ 2 in 2019 the routines keep different brands. The earlier value there (paired-swap PD p = 0.042) came from the mixed path and is superseded; the partial-sort rule applied to both observed and null graphs is reported as a sensitivity analysis |
| | | | | 0.0430; at top-100, ≥ 2 the node counts and observed PD agree under both rules. At top-111, ≥ 1 PD p = 0.0445 / 0.0225 (relabel / paired swap) under the brand-name rule against 0.0460 / 0.0245 under the partial sort; the battery separates nowhere under either rule. The primary observed graphs are the same under both rules; primary null draws changed where a draw tied at rank 60 (relabel PD p 0.261 → 0.259) | |
| Merge of five duplicate brand spellings (Brand spellings) | Implementation correction, after the earlier results were seen | 2026-09-27 | One label per brand, applied at the single data entry point before any count, ranking or basket is built | Distinct brands 138 → 133; brands offered per year 107, 111, 107, 98 → 106, 111, 104, 98. The primary node and edge sets are identical in every year; two 2020 edge weights rose by 1 (2020 weighted modularity 0.2060 → 0.2056). Relabel smallest Holm p 0.405 → 0.407 and paired-swap smallest Holm p 0.095 → 0.097, with the same PD p (0.259 | Found while checking a review item; the whole proprietary chain was rerun at the corrected entry point, and the earlier outputs are kept |
|  |  |  |  | and 0.118); the resolution sweep, the power designs and the descriptive analyses were rerun (values in the rows above) |  |
| Named-brand sensitivity analysis (Table S7) | Post hoc sensitivity analysis | Protocol written 2026-09-27, after the primary results and the descriptive metrics of the construction without non-brand labels were seen and before any sensitivity result | Reporting rule fixed before the run: if any Holm p < 0.05 under either full-year null, the abstract states the primary and the named-brand results side by side | Full year: under the paired-swap null the battery separates through mean degree (Holm p 0.030) and PD p = 0.037; under the relabel null the smallest Holm p is 0.187 and PD p = 0.098. April–December: all Holm p 1.000, PD p = 0.062. Planted-effect power on this construction: size 3/100 (battery and PD), battery MDE 0.3 (power 0.80 [0.708, 0.873]), PD MDE 0.8 | Not a pre-specified or primary test. A parity run with no label removed reproduced the primary results exactly before the sensitivity runs |
| Planted-effect power under further partitions of the anchor brands (Tables S3e, S3f) | Post hoc sensitivity analysis | Protocol written 2026-09-28, after the Design P results were seen and before this run | One rule for every outcome: report each partition's power at f = 0.3, the range, the median and the number of the 11 random partitions reaching 0.8; the category partition separately | Random partitions: 83–92 of 100 at f = 0.3 (median 85; at least 0.8 in 11 of 11), degree assortativity mean rejection rate 0.77; category partition 51, 90 and 99 of 100 at f = 0.2, 0.3 and 0.4 | None |
| Composition-only alternative with unit-year exposure labels (Table S3b(i)) | Post hoc sensitivity analysis | Protocol written 2026-09-28, after the Design S results were seen and before this run | Parity gate: the cell rule applied to all 2018–2021 sales reproduces every recorded label; then the S-high / S- | Parity passed (24,634 of 24,634 units, residency and four-cell labels). Applied to each buyer-year's | None |

| | | | low rule of Design S | own sales, the rule changes the residency label of 18 units (0.07%) and the cell of 176 (0.71%); battery 17/100 at a = 1 (S-low, as under the pooled labels), PD 7/100 | |
|---|---|---|---|---|---|
| Residual labels in the primary graphs (Table S7a note) | Descriptive, post hoc | Specified 2026-09-28 before the computation | None | Edges incident to "Other" and "No Brand" contribute −0.373 of the −1.759 difference in mean degree; the other edges −1.386 | None |

## Table S7. Named-brand sensitivity analysis (post hoc)

The primary construction uses the recorded brand labels, among which the residual labels "Other" and "No Brand" are nodes. This post hoc sensitivity analysis, specified after the primary results were seen and fixed in a written protocol before it ran (Table S6), removes six non-brand labels before top-60 ranking and otherwise repeats the primary construction and tests: the same units and group sizes, the same tie rule, threshold and graph builder, both nulls for the full-year comparison, the relabel null for April–December and its own planted-effect power curve. Holm corrects within each battery only; the analysis is not a second primary test. PD p-values are one-sided; metric p-values are two-sided.

**Table S7a. Construction.** Removed labels: Other, No Brand, K18 Gold, K24, Pt850 Platinum, Pt900 Platinum. Group sizes and units are those of the primary test; baskets emptied by the filter stay as units without brands.

| Quantity | 2019 | 2020 | April–December 2019 | April–December 2020 |
|---|---|---|---|---|
| Buyer-year (buyer-period) baskets | 14,235 | 10,399 | 11,078 | 6,775 |
| Transactions removed by the filter | 260 | 262 | 163 | 198 |
| Baskets emptied by the filter | 123 | 134 | 81 | 117 |
| Brands entering the top 60 after the filter | 2 | 2 | 2 | 2 |

| Nodes / edges | 36 / 115 | 31 / 72 | 33 / 92 | 24 / 53 |
|---|---|---|---|---|

Residual labels in the primary (recorded-label) graphs, descriptive: "Other" has degree 17 in 2019 and 9 in 2020, and "No Brand" 5 and 4 (the two are adjacent in 2020 only). The distinct edges incident to either label, 22 of 137 in 2019 and 12 of 84 in 2020, make up 2e_R/n = 1.100 and 0.727 of the mean degrees 6.850 and 5.091, so they contribute −0.373 of the −1.759 difference in mean degree and the edges among the other labels −1.386. This decomposition holds the primary graphs fixed; it is not the reranked named-brand construction below (Table S6).

**Table S7b. Full-year comparison, 2019 versus 2020 (R = 2,000 per null; seed 20260703).**

| Metric | 2019 | 2020 | Δ (2020 − 2019) | Relabel raw p | Relabel Holm p | Paired-swap raw p | Paired-swap Holm p |
|---|---|---|---|---|---|---|---|
| density | 0.183 | 0.155 | −0.028 | 0.311 | 0.933 | 0.302 | 0.907 |
| mean degree | 6.389 | 4.645 | −1.744 | 0.037 | 0.187 | 0.006 | 0.030 |
| clustering coefficient | 0.723 | 0.563 | −0.160 | 0.058 | 0.234 | 0.046 | 0.184 |
| modularity | 0.191 | 0.222 | 0.032 | 0.455 | 0.933 | 0.362 | 0.907 |
| degree assortativity | −0.575 | −0.590 | −0.014 | 0.723 | 0.933 | 0.722 | 0.907 |
| PD (diagnostic, one-sided) | – | – | 0.765 | 0.098 (195 of 2,000 draws ≥ observed) | – | 0.037 (73 of 2,000) | – |

Battery separation (any Holm p < 0.05): relabel null none; paired-swap null mean degree. Holm corrects within each battery only. Displayed Δ are computed at full precision and rounded once.

**Table S7c. Season-matched comparison, April–December 2019 against April–December 2020 (relabel null on buyer-period baskets, R = 2,000; seed 20261004).**

| Metric | Observed difference | Raw p (two-sided) | Holm p |
|---|---|---|---|
| density | 0.018 | 0.588 | 1.000 |
| mean degree | −1.159 | 0.295 | 1.000 |
| clustering coefficient | −0.018 | 0.896 | 1.000 |
| modularity | 0.029 | 0.615 | 1.000 |
| degree assortativity | 0.016 | 0.768 | 1.000 |
| PD (diagnostic) | 0.854 (observed) | 0.062 (one-sided; 123 of 2,000 draws ≥ observed) | – |

**Table S7d. Planted-effect power on the named-brand construction (Design P: 14,235 and 10,399 buyer-year baskets per replicate; redirection planted in the second group; 100 replicates per f; inner relabel null R = 999).**

| f | Replicates | Battery rejections | Battery defined | PD rejections | Density | Mean degree | Clustering | Modularity | Assortativity |
|---|---|---|---|---|---|---|---|---|---|

|  |  | (power) [95% CI] | share | (power) [95% CI] |  |  |  |  |  |
|---|---|---|---|---|---|---|---|---|---|
| 0 | 100 | 3/100 (0.03) [0.006, 0.085] | 1.00 | 3/100 (0.03) [0.006, 0.085] | 0.00 | 0.00 | 0.01 | 0.01 | 0.02 |
| 0.1 | 100 | 7/100 (0.07) [0.029, 0.139] | 1.00 | 10/100 (0.10) [0.049, 0.176] | 0.02 | 0.01 | 0.03 | 0.01 | 0.03 |
| 0.2 | 100 | 36/100 (0.36) [0.266, 0.462] | 1.00 | 16/100 (0.16) [0.094, 0.247] | 0.07 | 0.08 | 0.17 | 0.04 | 0.28 |
| 0.3 | 100 | 80/100 (0.80) [0.708, 0.873] | 1.00 | 18/100 (0.18) [0.110, 0.269] | 0.26 | 0.31 | 0.53 | 0.05 | 0.73 |
| 0.4 | 100 | 98/100 (0.98) [0.930, 0.998] | 1.00 | 29/100 (0.29) [0.204, 0.389] | 0.48 | 0.54 | 0.88 | 0.28 | 0.96 |
| 0.5 | 100 | 100/100 (1.00) [0.964, 1.000] | 1.00 | 41/100 (0.41) [0.313, 0.513] | 0.90 | 0.76 | 1.00 | 0.66 | 1.00 |
| 0.6 | 100 | 100/100 (1.00) [0.964, 1.000] | 1.00 | 49/100 (0.49) [0.389, 0.592] | 0.97 | 0.79 | 1.00 | 0.96 | 1.00 |
| 0.7 | 100 | 100/100 (1.00) [0.964, 1.000] | 1.00 | 63/100 (0.63) [0.528, 0.724] | 0.96 | 0.83 | 1.00 | 1.00 | 1.00 |
| 0.8 | 100 | 100/100 (1.00) [0.964, 1.000] | 1.00 | 83/100 (0.83) [0.742, 0.898] | 1.00 | 0.75 | 1.00 | 1.00 | 1.00 |
| 1.0 | 100 | 100/100 (1.00) [0.964, 1.000] | 1.00 | 98/100 (0.98) [0.930, 0.998] | 1.00 | 0.42 | 1.00 | 1.00 | 1.00 |

The f = 0 row is the empirical size of the whole procedure on this construction (battery 3/100, PD 3/100; the stop rule, 11 or more of 100, was not triggered). Battery MDE f = 0.3; PD MDE f = 0.8.

The anchor set is the pooled named-brand top 60, split once into two blocks of 30; the second group held on average 561 eligible baskets. The per-metric columns are Holm rejection rates.